\documentclass[%
 reprint,
 amsmath,amssymb,
 aps,
prl,
floatfix,
]{revtex4-1}

\usepackage{ifpdf}
\usepackage{amsmath} 
\usepackage{amssymb}
\usepackage{amsfonts}
\usepackage{braket}
\usepackage{color}
\usepackage[colorlinks=true,linkcolor=blue,citecolor=blue,breaklinks]{hyperref}
\usepackage{graphicx}
\usepackage{dcolumn}
\usepackage{bm}
\usepackage[normalem]{ulem}

\newcommand{\hc}{\text{H.c.}}
\newcommand{\la}{\langle}
\newcommand{\ra}{\rangle}
\newcommand{\tr}{\text{Tr}}

\newcommand{\Cph}{\mathcal{C}_{\text{ph}}}

\begin{document}


\title{Exact Quantum Many-Body Scar States in the Rydberg-Blockaded Atom Chain}
\author{Cheng-Ju Lin}
\affiliation{Department of Physics and Institute for Quantum Information and Matter, California Institute of Technology, Pasadena, California 91125, USA}
\author{Olexei I.~Motrunich}
\affiliation{Department of Physics and Institute for Quantum Information and Matter, California Institute of Technology, Pasadena, California 91125, USA}

\date{\today}

\begin{abstract}
A recent experiment in the Rydberg atom chain observed unusual oscillatory quench dynamics with a charge density wave initial state, and theoretical works identified a set of many-body ``scar states'' showing nonthermal behavior in the Hamiltonian as potentially responsible for the atypical dynamics.
In the same nonintegrable Hamiltonian, we discover several eigenstates at an \emph{infinite temperature} that can be represented exactly as matrix product states with a \emph{finite} bond dimension, for both periodic boundary conditions (two exact $E = 0$ states) and open boundary conditions (two $E = 0$ states and one each $E = \pm \sqrt{2}$).
This discovery explicitly demonstrates the violation of the strong eigenstate thermalization hypothesis in this model and uncovers exact quantum many-body scar states.
These states show signatures of translational symmetry breaking with period-2 bond-centered pattern, despite being in one dimension at an infinite temperature.
We show that the nearby many-body scar states can be well approximated as ``quasiparticle excitations" on top of our exact $E = 0$ scar states and propose a quasiparticle explanation of the strong oscillations observed in experiments.

\end{abstract}

\maketitle
{\it Introduction.--} Understanding quantum thermalization in isolated systems has attracted a lot of attention, due to both developments in cold atom experiments and fundamental theoretical interest.
The eigenstate thermalization hypothesis (ETH) has emerged as a paradigmatic mechanism for quantum thermalization~\cite{deutschQuantum1991, srednickiChaos1994}.
ETH postulates that a generic many-body system thermalizes at the level of individual eigenstates:
Eigenstates at the same energy density give the same expectation values of ``local-enough" observables.
The strong version of the ETH requires this on {\it every} eigenstate.
While an analytical proof is elusive, many numerical studies provided strong corroborations~\cite{rigolRelaxation2007, rigolThermalization2008, kimTesting2014, garrisonDoes2018}.
However, some systems showed atypical dynamics~\cite{banulsStrong2011, kormosRealtime2016} due to special low-energy states~\cite{linQuasiparticle2017a,jamesNonthermal2019,robinsonSignatures2018}.

A recent Rydberg cold atom experiment~\cite{bernienProbing2017} hinted at a new scenario, where the system exhibited atypical quench dynamics starting from a charge density wave (CDW) state at effective temperature $T = \infty$.
In contrast, a uniform initial state with the same energy density showed the expected thermalization behavior.
References~\cite{turnerWeak2018, turnerQuantum2018} proposed that this phenomenon is related to the presence of special eigenstates--{\it quantum many-body scar states}--which violate the ETH in the otherwise thermal spectrum, analogous to the nonergodic single-particle scar wavefunctions inside the chaotic single-particle spectrum~\cite{hellerBoundState1984}. 

Another nonintegrable system hosting nonthermal eigenstates is the Affleck-Lieb-Kennedy-Tasaki model~\cite{affleckRigorous1987}.
Reference~\cite{moudgalyaExact2018} constructed families of exact eigenstates in this model.
Using matrix product states (MPS), furthermore, Ref.~\cite{moudgalyaEntanglement2018} showed that these exact eigenstates with a finite energy density have logarithmic entanglement scaling in the subsystem size.
These papers thus provided an important analytical demonstration of exact scar states that violate the ETH~\cite{vafekEntanglement2017}.
Other works~\cite{shiraishiSystematic2017, moriThermalization2017} also proposed a special construction to embed nonthermal eigenstates into the many-body spectrum.

Remarkably, in the same Rydberg atom Hamiltonian studied in Refs.~\cite{bernienProbing2017, turnerWeak2018, turnerQuantum2018, hoPeriodic2019, khemaniSignatures2019}, we have discovered some exact scar states with a \emph{finite bond dimension} at energy density corresponding to $T = \infty$.
Our exact MPS description shows that these exact scar states have constant entanglement scaling and are, hence, even more ``nonthermal'' than the exact scar states at a finite energy density in Refs.~\cite{moudgalyaExact2018, moudgalyaEntanglement2018}.
Furthermore, these exact scar states break the lattice translation symmetry, despite being at $T = \infty$.
Thus, the strong ETH is violated in the Rydberg atom chain.
Using a ``single-mode approximation'' (SMA) and generaizing it to a ``multimode approximation" (MMA) on top of our exact scar states, we also find good approximations to nearby scar states, potentially relating the existence of other scar states to our exact states.

{\it Constrained Hilbert space and Hamiltonian.--}
Consider Rydberg atoms on a chain with $L$ sites, and denote $|0 \ra$ as the atomic ground state and $|1 \ra$ as the Rydberg excitation.
The Rydberg blockade prohibits states with $|\dots 1 1 \dots \ra$ on any two neighboring sites~\cite{bernienProbing2017}.
Despite the resulting non-tensor-product structure of the Hilbert space, one can still have the ETH concept~\cite{chandranEigenstate2016}.

The dynamics of this system is described by the so-called $PXP$ model:
\begin{equation}\label{eqn:Hc}
H = \sum_{j=2}^{L-1} P_{j-1} X_j P_{j+1} + H_1 + H_L ~,
\end{equation}
where $P = |0 \ra \la 0|$ is the projector to the Rydberg atom ground state and $X = |0 \ra \la 1| + |1 \ra \la 0|$ describes transitions between the ground and excited states.
(Previous works~\cite{sachdevMott2002, fendleyCompeting2004, lesanovskyManyBody2011, lesanovskyLiquid2012} studied low-energy states of related Hamiltonians.)
For periodic boundary conditions (PBC), we have $H_1 = P_L X_1 P_2$ and $H_L = P_{L-1} X_L P_1$, while for open boundary conditions (OBC), $H_1 = X_1 P_2$ and $H_L = P_{L-1} X_L$.
For PBC, the Hamiltonian has translation symmetry $T_x$ and inversion symmetry $I$, while for OBC, there is only inversion symmetry relative to the midpoint, $I: j \to L \!-\! j \!+\! 1$.
Furthermore, one can define ``particle-hole transformation'' $\Cph = \prod_j Z_j$, where $Z = |1 \ra \la 1| - |0 \ra \la  0|$.
This satisfies $\Cph H \Cph^{-1} = -H$, which guarantees that the spectrum is symmetric around zero energy; moreover, the intertwining of $\Cph$ with the inversion symmetry produces exponentially many zero-energy eigenstates~\cite{turnerQuantum2018, schecterManybody2018, turnerQuantum2018}.

The above Hamiltonian, despite its simple appearance, is not trivially solvable.
While its level-spacing statistics indicates its nonintegrability~\cite{turnerQuantum2018}, a recent work~\cite{khemaniSignatures2019} has suggested that it could be a deformation from some integrable Hamiltonian.

Inspired by Ref.~\cite{moudgalyaExact2018}, we inspected entanglement spectra of eigenstates of the $PXP$ model for OBC and discovered eigenstates at $E = \pm \sqrt{2}$ with a finite bond dimension.
We then reverse-engineered a simple MPS representation for these eigenstates and further identified two more exact eigenstates with $E = 0$ for OBC and two exact eigenstates at $E = 0$ for PBC.
Hence, these states analytically demonstrate that the $PXP$ Hamiltonian violates the strong ETH and are therefore {\it exact} quantum many-body scar states.

{\it Exact scar states for PBC.--}
These eigenstates exist for even $L$ (assumed throughout) and are expressed using MPSs.
We define $2 \times 3$ and $3 \times 2$ matrices
\begin{eqnarray}
B^0 &=&
\begin{pmatrix}
1 & 0 & 0 \\
0 & 1 & 0
\end{pmatrix} ~, ~~~~~
B^1 = \sqrt{2}
\begin{pmatrix}
0 & 0 & 0 \\
1 & 0 & 1
\end{pmatrix} ~, \\
C^0 &=&
\begin{pmatrix}
0 & -1 \\
1 & 0 \\
0 & 0
\end{pmatrix} ~, ~~~~~
C^1 = \sqrt{2}
\begin{pmatrix}
1 & 0 \\
0 & 0 \\
-1 & 0
\end{pmatrix} ~.
\end{eqnarray}
Two (unnormalized) exact scar states for PBC can be expressed as
\begin{eqnarray} \label{eqn:exactG}
|\Phi_1 \ra &=& \sum_{\{\sigma \}} \tr[B^{\sigma_1} C^{\sigma_2} \dots B^{\sigma_{L-1}} C^{\sigma_L}] |\sigma_1 \dots \sigma_L \ra ~, 
\end{eqnarray}
and $|\Phi_2 \ra = T_x |\Phi_1 \ra$, where $\sigma_j = 0$ or $1$.
The wavefunctions satisfy the constraints since $B^1 C^1 = 0_{2 \times 2}$ and $C^1 B^1 = 0_{3 \times 3}$.
In Supplemental Material~\cite{zotero-1083}, we prove $H |\Phi_i \ra = 0$.
Since these states are at $E = 0$, their effective temperature is $T = \infty$. 

The norm of the states is $\la \Phi_i | \Phi_i \ra = 3^{L_b} + 2 + (-1)^{L_b}$, where $L_b \equiv L/2$.
The two states are not orthogonal and have overlap $\la \Phi_1 | \Phi_2 \ra = 2 [(\sqrt{2} - 1)^{L_b} + (-1)^{L_b} (\sqrt{2} + 1)^{L_b}]$; however, they are linearly independent for $L_b > 3$ [for $L_b \leq 3$, we happen to have $|\Phi_2 \ra = (-1)^{L_b} |\Phi_1 \ra$].
For $L_b > 3$, the states $|\Phi_{1,2} \ra$ in fact break the translation symmetry $T_x$, while by construction they are invariant under $T_x^2$.
One can form degenerate states $|\Phi_{K=0/\pi} \ra = |\Phi_1 \ra \pm |\Phi_2 \ra$ that carry definite momenta $0$ and $\pi$, which can be viewed as a finite-size signature of the $T_x$ breaking that appears in the thermodynamic limit.

Let us examine properties of the state $|\Phi_1 \ra$ (properties of $|\Phi_2 \ra$ simply follow).
First, the breaking of $T_x$ in this state cannot be detected by any one-site observable, since the one-site reduced density matrices are the same for all sites, $\rho^{\text{one-site}} = \frac{2}{3} |0 \ra \la 0| + \frac{1}{3} |1 \ra \la 1|$ in the thermodynamic limit~\cite{zotero-1083}.
In particular, for the Rydberg excitation number $n_j = |1 \ra \la 1|$, we have $\la \Phi_1 | n_j |\Phi_1 \ra / \la \Phi_1 | \Phi_1 \ra = \frac{1}{3}$.
This violates the ETH, since, at $T = \infty$, the Gibbs ensemble predicts $\la n_j \ra_{T = \infty} = (1 + \phi^2)^{-1} \approx 0.2764$, where $\phi = (1 + \sqrt{5})/2$ is the golden ratio.

On the other hand, two-site observables can detect the $T_x$ breaking, as can be seen from the corresponding reduced density matrices for subsystems $[1,2]$ and $[2,3]$ in the $|\Phi_1 \ra$ state:
\begin{align}\label{eqn:red_mat_12}
\rho^{\text{two-site}}_{[1,2]} &= \frac{1}{3} (|0 0 \ra \la 0 0| + |0 1 \ra \la 0 1| + |1 0 \ra \la 1 0|) ~, \\
\label{eqn:red_mat_23}
\rho^{\text{two-site}}_{[2,3]} &= \frac{1}{3} (|0 0 \ra \la 0 0| + |0 1 \ra \la 0 1| + |1 0 \ra \la 1 0|) \nonumber \\
& - \frac{1}{9} (|0 1 \ra \la 1 0| + |1 0 \ra \la 0 1|) ~.
\end{align}
In particular, we see that $|0_j 1_{j+1} \ra \la 1_j 0_{j+1}| + \hc$ has expectation value $0$ for $j$ odd and $-2/9$ for $j$ even.

We also list the symmetry properties of these exact scar states (see Ref.~\cite{zotero-1083} for the proof).
For $L$ even, the inversion $I$ defined earlier is relative to a bond center and is not broken.
We find $I |\Phi_1 \ra = (-1)^{L_b} |\Phi_1 \ra$.
For $|\Phi_2 \ra$, note that since $I T_x = T_x^{-1} I$ and $T_x^2 |\Phi_i \ra = |\Phi_i \ra$, we obtain also $I |\Phi_2 \ra = (-1)^{L_b} |\Phi_2 \ra$.
While $\Cph$ is not a symmetry of $H$, our states are, in fact, eigenstates of $\Cph$.
We have $\Cph |\Phi_i \ra = (-1)^{L_b} |\Phi_i \ra$ for both $i = 1, 2$.

{\it Exact scar states for OBC.-}
We also found exact scar states for OBC with the same bulk MPSs.
Defining ``boundary vectors'' $v_1 = (1, 1)^T$ and $v_2 = (1, -1)^T$, we can write four exact scar states
\begin{equation}
|\Gamma_{\alpha, \beta} \ra = \sum_{\{\sigma\}} v_\alpha^T B^{\sigma_1} C^{\sigma_2} \dots B^{\sigma_{L-1}} C^{\sigma_L} v_\beta |\sigma_1 \dots \sigma_L \ra ~,
\label{eqn:GammaOBC}
\end{equation}
where $\alpha, \beta \in \{1, 2\}$.
The eigenenergies are $E = 0$ for $|\Gamma_{\alpha,\alpha} \ra$, $E = \sqrt{2}$ for $|\Gamma_{1,2} \ra$, and $E = -\sqrt{2}$ for $|\Gamma_{2,1}\ra$, see Ref.~\cite{zotero-1083}.

\begin{figure}
    \includegraphics[width=1.0\columnwidth]{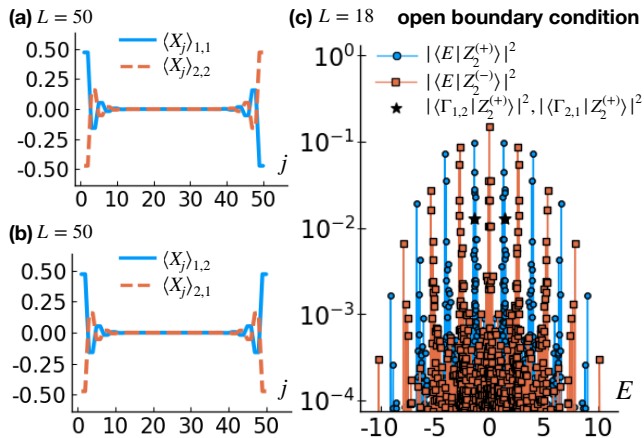}
    \caption{ 
    (a)(b) Energy density profiles $\la X_j \ra_{\alpha,\beta}$ in the four exact eigenstates $|\Gamma_{\alpha,\beta} \ra$ in the OBC system of size $L = 50$.
    (c)Towers of the $Z_2$ scar states for OBC found in ED.
    The positions of the exact scar states $|\Gamma_{1,2} \ra$ and $|\Gamma_{2,1} \ra$ are marked with stars.
    }
    \label{fig:OBCenergy}
\end{figure}

It is interesting to examine the energy density profiles. 
Figures~\ref{fig:OBCenergy}(a) and \ref{fig:OBCenergy}(b) show $\la X_j \ra_{\alpha,\beta} \equiv \la \Gamma_{\alpha,\beta} | X_j |\Gamma_{\alpha,\beta} \ra / \la \Gamma_{\alpha,\beta} | \Gamma_{\alpha,\beta} \ra$ in each state~\cite{zotero-1083}.
We can see that there are localized ``energy lumps'' at the edges of the chain.
The profiles decay exponentially into the bulk with decay length $2 \ln(3)$.
The integrated energy over each lump is $\sqrt{2}/2$ or $-\sqrt{2}/2$ depending on the termination, which can be thought as representing different ``edge states.''

The symmetry properties of $|\Gamma_{\alpha,\beta} \ra$ can be derived in a similar fashion as for PBC~\cite{zotero-1083}.
In particular, we have $I |\Gamma_{1,2} \ra = (-1)^{L_b-1} |\Gamma_{1,2} \ra$ and $I |\Gamma_{2,1} \ra = (-1)^{L_b-1} |\Gamma_{2,1} \ra$; while $I| \Gamma_{1,1} \ra = (-1)^{L_b} |\Gamma_{2,2} \ra$ and $I |\Gamma_{2,2} \ra = (-1)^{L_b} |\Gamma_{1,1} \ra$.
As for the particle-hole transformation, we obtain $\Cph |\Gamma_{1,2} \ra = (-1)^{L_b} |\Gamma_{2,1} \ra$ and $\Cph |\Gamma_{1,1} \ra = (-1)^{L_b} |\Gamma_{2,2} \ra$.
The fact that $|\Gamma_{1,2} \ra$ and $|\Gamma_{2,1} \ra$ are eigenstates of $I$ means that they can be nondegenerate, which is what we found in exact diagonalization (ED). 
As expected, these $E = \pm \sqrt{2}$ scar states are related by $\Cph$. 
Since they are nondegenerate, their finite bond dimensions are not related to the exponential degeneracy of the $E = 0$ sector.
Their existence again demonstrates the violation of the ETH, even without worrying about potential subtleties in the degenerate space~\cite{schecterManybody2018}.

We can also calculate entanglement in $|\Gamma_{\alpha,\beta}\ra$ for any cut and system size~\cite{ciracEntanglement2011, moudgalyaEntanglement2018}.
In the thermodynamic limit, across a cut between $C_{2b}$ and $B_{2b+1}$ (bond-dimension $D=2$ cut), we find~\cite{zotero-1083} the squared Schmidt values $1/2$ and $1/2$, which gives the von Neumann entanglement entropy $S_{\text{vN}}^{\text{OBC}, D=2} = \ln 2$.
Cutting instead across $B_{2b+1}$ and $C_{2b+2}$ ($D=3$), the squared Schmidt values are $2/3$, $1/6$ and $1/6$, and $S_{\text{vN}}^{\text{OBC}, D=3} = -\frac{2}{3} \ln(\frac{2}{3}) - \frac{1}{3} \ln(\frac{1}{6}) \approx 0.868$.

For the states $|\Phi_i \ra$ in PBC and a large subregion, there are two entanglement cuts, and the entanglement entropy will be the sum of the OBC entropies associated with each cut (and will remain finite in the thermodynamic limit).
We can then predict that for the states $|\Phi_{K=0/\pi} \ra$, the entanglement entropy will be $S_{\text{vN}}^{\text{PBC}} = S_{\text{vN}}^{\text{OBC}, D=2} + S_{\text{vN}}^{\text{OBC}, D=3} + \ln 2 \approx 2.254$.

\begin{figure}
    \includegraphics[width=1.0\columnwidth]{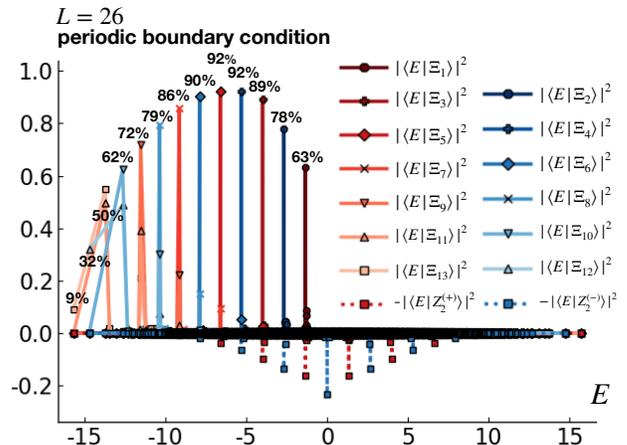}
    \caption{Overlaps of the SMA and MMA  wavefunctions with the eigenstates in the PBC system with $L = 26$.
    We also list the overlaps with the primary $Z_2$ scar states. 
    The $Z_2$ scar states are identified through the overlaps with the $|Z_2^{(+)} \ra$ or $|Z_2^{(-)} \ra$ states (for more clarity, we show negatives of these overlaps).
    }
    \label{fig:QpAnsatz}
\end{figure}

{\it Possible relation to $Z_2$ scar states.-} Turner {\it et.~al.} \cite{turnerWeak2018, turnerQuantum2018} focused on the $PXP$ model with PBC and identified a set of quantum many-body scar states (called $Z_2$ scar states) through the overlap of eigenstates $|E \ra$ with the CDW states $|Z_2 \ra = |1 0 \dots 1 0 \ra$ or $|Z_2' \ra = |0 1 \dots 0 1\ra$.
The most prominent such scar states have the largest overlap and the smallest entanglement entropy compared to nearby states, but there are also ``bands" (or ``towers") of weaker scar states close to each primary one.
The consecutive primary scar states have an almost equal energy separation of $\approx 1.33$.
The scar states and this frequency were proposed to be responsible for the strong oscillations observed in quenches from the $|Z_2 \ra$ state.

It is convenient to consider states $|Z_2^{(\pm)} \ra = (|Z_2 \ra \pm |Z_2' \ra)/\sqrt{2}$, which have inversion quantum numbers $I \!=\! 1$ and $I \!=\! -1$ and carry momenta $K \!=\! 0$ and $K \!=\! \pi$ respectively, if in PBC.
For $L_b$ even, the $Z_2$ scar states at energy $E \approx 0$ are found to have $I \!=\! 1$ (and $K \!=\! 0$ in PBC), while for $L_b$ odd they have $I \!=\! -1$ (and $K \!=\! \pi$).
For a fixed $L_b$, $I$ (and $K$ in PBC) alternate between these values when going from one primary scar state to the next (and are the same within the band of weaker scar states associated with each primary state).
This is illustrated in Figs.~\ref{fig:OBCenergy}(c) and~\ref{fig:QpAnsatz}.

Turner~{\it et.~al.}~\cite{turnerWeak2018, turnerQuantum2018} proposed to approximate the primary scar states using ``forward scattering approximation" (FSA) starting from the $Z_2$ state.
We propose an alternative picture starting from our exact $E = 0$ states.

First, we note that our exact $E = 0$ scar states are, in fact, representative of the nearby scar states.
For instance, at $L = 26$, the nearby $E \approx \pm 1.34$ scar states have average Rydberg excitation number $\la E | n_j| E \ra \approx 0.3476$ while $\la \Phi_{K=\pi} | n_j | \Phi_{K=\pi} \ra \approx 0.3355$.
Second, we note that for OBC, the exact scar states $|\Gamma_{1,2} \ra$ and $|\Gamma_{2,1} \ra$, while not being the primary $Z_2$ scar states, belong to the first non-zero-energy towers of scar states, as shown in Fig.~\ref{fig:OBCenergy}(c).
Furthermore, we can understand these exact $E = \pm \sqrt{2}$ scar states as ``edge excitations'' on top of the $E = 0$ states $|\Gamma_{\alpha,\alpha} \ra$ (see Ref.~\cite{zotero-1083}).
We therefore conjecture that for the PBC system as well, the nearby scar states can be understood as quasiparticle excitations on top of the ``vacuum" $|\Phi_i \ra$.

Motivated by these observations, we construct variational wavefunctions using SMA~\cite{haegemanVariational2012, haegemanElementary2013} and generalize it to MMA on top of our exact $|\Phi_i \ra$ states and aimed to capture the nearby scar states.
We start with the following SMA wavefunction $|\Xi_1 \ra  = [|M_1 \ra - (-1)^{L_b} T_x |M_1 \ra]/\xi_1$, where
\begin{align}{\label{eqn:Mtrial}}
|M_1 \ra \! =\! \sum_{\{\sigma\}}\sum_{b=1}^{L_b} \tr[B^{\sigma_1} C^{\sigma_2} \!\! \dots \! M^{\sigma_{2b-1} \sigma_{2b}} \!\! \dots \! C^{\sigma_L}] |\sigma_1 \! \dots \! \sigma_L \ra, 
\end{align}
and $\xi_1$ provides normalization $\la \Xi_1 | \Xi_1 \ra = 1$.
The matrices
\begin{align*}
M^{00} = \begin{pmatrix}
1 & 0 \\
0 & 1
\end{pmatrix} ~,~
M^{01} = \begin{pmatrix}
\mu_1 & 0 \\
\mu_2 & 0
\end{pmatrix} ~,~
M^{10} = \begin{pmatrix}
0 & 0 \\
-\mu_2 & \mu_1
\end{pmatrix}
\end{align*}
are chosen such that the wavefunction satisfies the Rydberg-blockaded constraint and $I |M_1 \ra = (-1)^{L_b-1} |M_1 \ra$, hence $I |\Xi_1 \ra = (-1)^{L_b-1} |\Xi_1 \ra$ (see Ref.~\cite{zotero-1083}).
We have also chosen the translation quantum number of $|\Xi_1 \ra$ to be $(-1)^{L_b-1}$, which matches the symmetry sector of the first $E \neq 0$ scar state overlapping with the $Z_2$ CDW.
To make $|\Xi_1 \ra$ as close to an eigenstate as possible, we minimize the energy variance $\sigma_H^2(\mu_1, \mu_2) = \la \Xi_1 | H^2 |\Xi_1 \ra - \la \Xi_1 | H | \Xi_1 \ra^2$ at fixed $L$.
At $L = 26$, we find optimal parameters $\mu_1 = -1.0876$ and $\mu_2 = -0.6344$, which give $\sigma_H^2 = 0.0263$ and the average energy $\la \Xi_1 |H |\Xi_1 \ra = -1.3147$.
Remarkably, the optimized state has over $63\%$ overlap with the primary $Z_2$ scar state at $E \approx -1.3386$ found in ED, as shown in Fig.~\ref{fig:QpAnsatz}.
It is easy to check that $\mu_1^\prime = -\mu_1, \mu_2^\prime = \mu_2$ gives $|\Xi_1^\prime \ra = (-1)^{L_b - 1} \Cph |\Xi_1 \ra$, which captures the scar state with $E \approx 1.3386$.

To capture other primary scar states and support our picture of quasiparticle excitations, we examine the following MMA wavefunctions $|\Xi_n \ra = [|M_n \ra + (-1)^{L_b + n} T_x |M_n \ra]/\xi_n$, where
\begin{align}
    |M_n \ra = & \sum_{\{\sigma\}}\sideset{}{'}\sum_{b_1, \dots, b_n = 1}^{L_b} \tr[B^{\sigma_1} C^{\sigma_2} \dots M^{\sigma_{2b_1-1} \sigma_{2b_1}} \dots \notag \\
    & \dots M^{\sigma_{2b_n-1} \sigma_{2b_n}} \dots B^{\sigma_{L-1}} C^{\sigma_L}] |\sigma_1 \dots \sigma_L \ra ~,
\end{align}
and the summation is constrained to have all $b_i$ distinct and $\xi_n$ is the normalization factor.
Such an $|M_n \ra$ describes some $n$-particle scattering state and is the most primitive construction where we simply try hard-core exclusion of the particles.
For simplicity, we will take $M$ from the optimization of $|\Xi_1 \ra$.
Moreover, $|\Xi_n\ra$ has quantum numbers $T_x=(-1)^{L_b+n}$ and $I=(-1)^{L_b+n}$, matching the symmetry structure of the $Z_2$ scar states.
Unexpectedly, Fig.~\ref{fig:QpAnsatz} shows that the overlaps of such simplest MMA wavefunctions and the primary scar states become better with more quasiparticles, up to about $n \approx L_b/2$, while for larger $n$ the overlaps start to decrease.
The poorer performance for $n > L_b/2$ is not surprising:  For example, for $n = L_b$, the state $|M_{L_b}\ra=
\bigotimes_{b=1}^{L_b}|0\ra_{2b-1}(|0\ra+\mu_1|1\ra)_{2b} + \bigotimes_{b=1}^{L_b}(|0\ra+\mu_1|1\ra)_{2b-1}|0\ra_{2b}-\bigotimes_{b=1}^{L_b}|0\ra|0\ra$, therefore $|\Xi_{L_b} \ra \sim |M_{L_b}\ra$ but has spontaneous $T_x$ symmetry breaking and is only a crude approximation to the true nondegenerate fully symmetric ground state.
Our MMA states with $n$ close to $L_b$ are similarly expected to be only crude approximations to the actual primary scar states and are seen to be spread over several nearby scar states.
On the other hand, the performance of the states with $n < L_b/2$ is truly remarkable.
Typically, when adding more quasiparticles without further optimizations, such MMA states become worse with the number of particles added, while our MMA have better overlaps with the primary scar states.
Furthermore, our MMA states perform better than the FSA states for $2 \leq n \simeq L_b/2$.
For reference, at $L = 26$, the FSA states have overlap $69\%$ with the scar states $E \approx \pm 1.33$ and $68\%-72\%$, overlaps on the consecutive primary scar states respectively~\cite{zotero-1083}. 
This suggests that our exact eigenstates at $E = 0$ provide a better starting point for understanding the scar states in the $PXP$ model.

Let us further discuss these results.
$|\Xi_1 \ra$ and $|\Xi_1^\prime \ra \sim \Cph |\Xi_1 \ra$ can be viewed as representing ``elementary quasiparticles" with energies $\epsilon_- \approx -1.31$ and $\epsilon_+ = -\epsilon_-$; these particles also carry inversion quantum number $-1$.
It is then natural to expect strong oscillations with frequency $\epsilon_+$ in observables that flip the inversion quantum number.
(Observables in experiment and numerics that do not flip $I$ will show frequency $2\epsilon_+$.)
Indeed, even though the overlaps of the $Z_2$ initial state with the primary scar states decrease exponentially with the system size, the ``quasiparticle creation operators'' can also act on many more states, always ``adding'' roughly $\epsilon_\pm$.
This argument resembles the quasiparticle explanation~\cite{linQuasiparticle2017a} of strong oscillations in the ``weak thermalization" regime in Ref.~\cite{banulsStrong2011}, where the initial state happened to be near the ground state.
The differences here are that the initial $Z_2$ state is at $T = \infty$ but is ``close'' to our special eigenstates $|\Phi_i \ra$, and that the quasiparticles here can carry both positive and negative energies.

By the repeated application of the SMA construction that gave us the $|\Xi_1 \ra$ and $|\Xi_1^\prime \ra$ states, we also expect additional states with energies $E \approx (n_+ - n_-) \epsilon_+$, $n_+, n_- \in \mathbb{N}$.
We have demonstrated the $(n_+, n_-) = (0, n)$ branch explicitly in Fig.~\ref{fig:QpAnsatz}.
Interestingly, the same energy $m \epsilon_+$ can be obtained in multiple ways, which may explain the bands of weaker scar states near the primary states.

Finally, we note that the presented simple ``bond-dimension-2" SMA wavefunctions cover cases where we replace one $B$ or one $C$ with an ``excitation,'' or ``excite'' two consecutive $B_{2b-1}, C_{2b}$ matrices.
One can also consider exciting two consecutive $C_{2b}, B_{2b+1}$ matrices, which would lead to new ``bond-dimension-3'' SMA wavefunctions with more variational parameters and the corresponding MMA wavefunctions.
Our study shows that they can capture the primary $Z_2$ scar states with even higher fidelity~\cite{zotero-1083}, but since the improvement is only quantitative, we presented the simpler bond-dimension-2 SMA.

{\it Conclusions.--}
We discovered exact scar states in the Rydberg-blockaded atom chain at $T = \infty$ that explicitly violate the strong ETH and have constant entanglement scaling in the subsystem size.
Our exact states show translation symmetry breaking, which implies twofold degeneracy for PBC.
The exact scar states for OBC have the same bulk as for PBC and can have different edge terminations leading to different eigenenergies, including nondegenerate energies.

By constructing quasiparticles on top of the exact scar states, we capture the primary $Z_2$ scar states with high fidelity.
Systematic improvements for capturing the primary scar states, as well as study bands of weaker scar states are therefore warranted.
For example, even for the SMA, is there a convergent construction that reproduces the first primary $Z_2$ scar state and proves its ETH-violating properties?
It is also interesting to understand the pattern of scar states in the $PXP$ model more generally and how it compares with other instances of exact scar states~\cite{moudgalyaExact2018, moudgalyaEntanglement2018, shiraishiSystematic2017}.
Studying additional models with exact scar states and their stability to perturbations would be beneficial for both of these questions. 
We leave such explorations for future work.

\begin{acknowledgments}
We thank V. Albert, M. Endres, B. Roberts, B. Timar, and C. White for valuable discussions.
This work was supported by National Science Foundation (NSF) through Grant No. DMR-1619696, and also by the Institute for Quantum Information and Matter, an NSF Physics Frontiers Center, with support of the Gordon and Betty Moore Foundation. 
\end{acknowledgments}

%


\end{document}



\title{Supplemental Material: Exact Quantum Many-body Scar States in the Rydberg-blockaded Atom Chain}
\author{Cheng-Ju Lin}
\affiliation{Department of Physics and Institute for Quantum Information and Matter, California Institute of Technology, Pasadena, CA 91125, USA}
\author{Olexei I.~Motrunich}
\affiliation{Department of Physics and Institute for Quantum Information and Matter, California Institute of Technology, Pasadena, CA 91125, USA}




\date{\today}

\maketitle

\onecolumngrid
\section{Proof of exact scar states in PBC}
\label{app:proofPBC}
Here we prove that $H |\Phi_1 \ra = 0$. 
To prove this, it is easier to work in the blocked reformulation of the Hamiltonian:
We block two sites $2b-1, 2b$ into one ``block-site", with allowed block states $(00)$, $(10)$, and $(01)$ denotes as $O$, $L$, and $R$ respectively; the Rydberg constraint further disallows configurations with $RL$ on consecutive blocks.
The number of blocks is $L_b = L/2$, and recall that throughout we assume that $L$ is even.
In the blocked representation, the Hamiltonian can be written as a sum of two-body terms:
\begin{align}
H = \sum_{b = 1}^{L_b} h_{b, b+1} ~, \quad
h_{b, b+1} = (|R \ra \la O| + |O \ra \la R|)_b \otimes (I - |L \ra \la L|)_{b+1} 
+ (I - |R \ra \la R|)_b \otimes (|L \ra \la O| + |O \ra \la L|)_{b + 1} ~.
\label{eqn:Hblocked}
\end{align}

The state $|\Phi_1 \ra$ can be written in the blocked representation as an MPS of bond dimension 2, namely
\begin{align}
|\Phi_1 \ra =\sum_{\{s\}} \tr [A^{s_1} \dots A^{s_{L_b}}] |s_1 \dots s_{L_b} \ra ~,
\end{align}
where we have introduced blocked matrices $A^{(\sigma_1 \sigma_2)} = B^{\sigma_1} C^{\sigma_2}$.
Explicitly,
\begin{align}
A^O = \begin{pmatrix}
0 & -1 \\
1 & 0
\end{pmatrix} ~, \quad
A^R = \begin{pmatrix}
\sqrt{2} & 0 \\
0 & 0 
\end{pmatrix} ~, \quad
A^L = \begin{pmatrix}
0 & 0 \\
0 & -\sqrt{2}
\end{pmatrix} ~.
\label{eq:Amatr}
\end{align}
One can easily check that $A^R A^L = 0$, so the state satisfies the Rydberg constraint between the blocks.
Interestingly, these matrices also satisfy $A^L A^R = 0$, so this state also disallows $LR$ on consecutive blocks.

We first examine how the genuinely two-body part of the Hamiltonian term $h_{b,b+1}$, namely $h^{(2)}_{b,b+1} \equiv -(|R \ra \la O| + |O \ra \la R|)_b \otimes (|L \ra \la L|)_{b+1} - (|R \ra \la R|)_b \otimes (|L \ra \la O| + |O \ra \la L|)_{b+1}$ operates on $|\Phi_1 \ra$.
The special property $A^R A^L = 0$ leaves us only the part $-(|R \ra \la O|)_b \otimes (|L \ra \la L|)_{b+1} - (|R \ra \la R|)_b \otimes (|L \ra \la O|)_{b+1}$.
It is easy to check that the matrices also satisfy $A^O A^L + A^R A^O = 0$, and hence we conclude that $h^{(2)}_{b,b+1} |\Phi_1 \ra = 0$.

We now collect the one-body parts of the Hamiltonian and after convenient grouping obtain:
\begin{equation}
    H' = \sum_{b=1}^{L_b} \Big(|R \ra \la O| + |O \ra \la R| + |L \ra \la O| + |O \ra \la L| \Big)_b ~.
\label{eq:Hprime}
\end{equation}
Consider action of a term associated with block $b$ on $|\Phi_1 \ra$:
\begin{equation}
\Big(|R \ra \la O| + |O \ra \la R| + |L \ra \la O| + |O \ra \la L| \Big)_b |\Phi_1 \ra = \sum_{\{s\}} \tr[A^{s_1} \dots F^{s_b} \dots A^{s_{L_b}}] |s_1 \dots s_{L_b} \ra ~,
\label{eqn:onebodyH_Phi1}
\end{equation}
where
\begin{equation}
F^O = \begin{pmatrix}
\sqrt{2} & 0 \\
0 & -\sqrt{2}
\end{pmatrix} ~, \quad
F^R = F^L = \begin{pmatrix}
0 & -1 \\
1 & 0
\end{pmatrix} ~.
\label{eq:Fmatr}
\end{equation}
Therefore, we have 
\begin{equation}\label{eqn:HPhi_1}
H |\Phi_1 \ra =  H' |\Phi_1 \ra = \sum_{b=1}^{L_b} \sum_{\{s\}} \tr[A^{s_1} \dots F^{s_b} \dots A^{s_{L_b}}] |s_1 \dots s_{L_b} \ra ~.
\end{equation}
It is easy to verify that $F^s = X A^s - A^s X$, where $X = \frac{1}{\sqrt{2}} \sigma_x$.
Substituting this in Eq.~(\ref{eqn:HPhi_1}), we therefore see that $H |\Phi_1 \ra = 0$.
The fact that $|\Phi_2 \ra$ is also an eigenstate follows from the translational invariance of the Hamiltonian: $H |\Phi_2 \ra = H T_x |\Phi_1 \ra = T_x H |\Phi_1 \ra = 0$.

It is instructive to see an alternative proof how the sum of the one-body terms $H'$ annihilates $\Phi_1$ (which will be also useful later for developing  intuition about our single-mode approximation constructed on top of $|\Phi_1 \ra$).
To this end, we first transform to the basis diagonalizing the one-body terms.
On each block-site, the eigenvalues are $\sqrt{2}$, $-\sqrt{2}$, and $0$, and the corresponding eigenvectors are
\begin{align}
|\bm{+} \ra_b \equiv \frac{1}{2}\left(|R \ra + |L \ra + \sqrt{2} |O \ra \right)_b ~, \quad 
|\bm{-} \ra_b \equiv \frac{1}{2}\left(|R \ra + |L \ra - \sqrt{2} |O \ra \right)_b ~, \quad 
|\bm{0} \ra_b \equiv \frac{1}{\sqrt{2}} \left(|R \ra - |L \ra \right)_b ~.
\label{eq:+-0basis}
\end{align}
Hence we have
\begin{equation}
H' = \sqrt{2} \sum_{b=1}^{L_b} \Big(|\bm{+} \ra \la \bm{+}| - |\bm{-} \ra \la \bm{-}| \Big)_b = \sqrt{2} (N_{\bm{+}} - N_{\bm{-}}) ~,
\label{eq:Hprime+-0}
\end{equation}
where $N_{\bm{\pm}} \equiv \sum_{b=1}^{L_b} (|\bm{\pm} \ra \la \bm{\pm}|)_b$ simply count numbers of $\bm{+}$ vs $\bm{-}$ states on the block-site lattice.

In the new basis, we can write $|\Phi_1 \ra$ as an MPS with matrices $A^{\bm{\pm}} = V \frac{1}{2}\left(A^R + A^L \pm \sqrt{2} A^O \right) V^{-1}$, $A^{\bm{0}} = V \frac{1}{\sqrt{2}} \left(A^R - A^L \right) V^{-1}$, where it was also convenient 
to perform an additional gauge transformation with $V = \frac{1}{\sqrt{2}} \begin{pmatrix} 1 & 1 \\ 1 & -1 \end{pmatrix}$.
The resulting matrices are
\begin{align}
A^{\bm{+}} = \begin{pmatrix}
0 & \sqrt{2} \\
0 & 0
\end{pmatrix} ~, \quad
A^{\bm{-}} = \begin{pmatrix}
0 & 0 \\
\sqrt{2} & 0 
\end{pmatrix} ~, \quad
A^{\bm{0}} = \begin{pmatrix}
1 & 0 \\
0 & 1
\end{pmatrix} ~.
\label{eq:Amatr+-0}
\end{align}
Using properties $(A^{\bm{+}})^2 = (A^{\bm{-}})^2 = 0$, $A^{\bm{+}} A^{\bm{-}} = \text{diag}(2, 0)$, and the fact that $A^{\bm{0}}$ is identity matrix, it is now easy to expand $|\Phi_1 \ra$ in the $\bm{\pm}, \bm{0}$ basis.
In particular, we see that for basis vectors with non-trivial contributions to $|\Phi_1 \ra$, each $\bm{+}$ must be followed by $\bm{-}$, with possibly intervening $\bm{0}$'s in any number.
This immediately implies that in each such basis vector we have $N_+ = N_-$; hence, $|\Phi_1 \ra$ is indeed annihilated by the sum of the one-body terms $H'$.

\section{Relation between $|\Phi_1 \ra$ and AKLT state}
It is interesting to note that there is a precise relation between our exact eigenstate $|\Phi_1 \ra$ in the blocked representation and the celebrated AKLT state in a spin-1 chain.
Specifically, we can perform the following gauge transformation $\tr[A^{s_1} A^{s_2} A^{s_3} A^{s_4} \dots] = \tr[A^{s_1} U U^{-1} A^{s_2} A^{s_3} U U^{-1} A^{s_4} \dots] = \tr[(A^\prime)^{s_1} (A^{\prime\prime})^{s_2} (A^\prime)^{s_3} (A^{\prime\prime})^{s_4} \dots]$ with $U = \sigma^x$ Pauli matrix and $(A^\prime)^s = A^s U$, $(A^{\prime\prime})^s = U^{-1} A^s$.
The matrices $(A^\prime)^s$ are precisely the matrices used in an MPS representation of the AKLT state with identification $s = O, R, L$ as $S_z = 0, 1, -1$ in the spin-1 chain, while the matrices $(A^{\prime\prime})^s$ become the same as $(A^\prime)^s$ after a unitary transformation on the physical states that interchanges $L$ and $R$ states.
Unfortunately, the Hamiltonians in the Rydberg problem and in the AKLT problem appear to be drastically different.
Most notably, the Rydberg Hamiltonian has a nontrivial translation symmetry $T_x$ by one Rydberg atom, while the AKLT Hamiltonian ``knows'' only about $T_x^2$ which is the simple translation symmetry by one block in the blocked variables.
Also, the AKLT Hamiltonian has continuous spin rotation symmetry and is a sum of local terms that individually annihilate the AKLT state, which is not the case for the Rydberg Hamiltonian and our exact eigenstate. 
So far, we have not been able to utilize knowledge about the AKLT Hamiltonian in the Rydberg problem.

\section{Proof of exact eigenstates in OBC}
\label{app:proofOBC}
Here we prove that $|\Gamma_{\alpha,\beta} \ra$ defined in Eq.~(7) in the main text are eigenstates in OBC.
In this case, the Hamiltonian in the blocked form is
\begin{equation}
H = \sum_{b=1}^{L_b-1} h_{b,b+1} + h_{\text{left}} + h_{\text{right}} ~, \quad
h_{\text{left}} = (|L \ra \la O| + |O \ra \la L|)_{b=1} ~, \quad
h_{\text{right}} = (|R \ra \la O| + |O \ra \la R|)_{b=L_b} ~,
\end{equation}
with the ``bulk'' $h_{b,b+1}$ given in Eq.~(\ref{eqn:Hblocked}).

We can write the states in the blocked representation as
$|\Gamma_{\alpha,\beta} \ra = \sum_{\{s\}} v_{\alpha}^T A^{s_1} \dots A^{s_{L_b}} v_{\beta} |s_1 \dots s_{L_b} \ra$.
Similar to the case in PBC, the genuinely two-body part of $h_{b,b+1}$ annihilates these states, $h^{(2)}_{b,b+1} |\Gamma_{\alpha,\beta} \ra = 0$.
We therefore have
\begin{equation}
H |\Gamma_{\alpha,\beta} \ra = H' |\Gamma_{\alpha,\beta} \ra = \sum_{b=1}^{L_b} \sum_{\{s\}} v^T_\alpha A^{s_1} \dots F^{s_b} \dots A^{s_{L_b}} v_\beta |s_1 \dots s_{L_b} \ra ~,
\label{eqn:HGammaOBC}
\end{equation}
where $H'$ is the sum of one-body terms defined in Eq.~(\ref{eq:Hprime}), and matrices $F^s$ are defined in Eq.~(\ref{eq:Fmatr}).
We can again substitute
$F^s = X A^s - A^s X$, with $X = \frac{1}{\sqrt{2}} \sigma^x$, and obtain
\begin{equation}
H |\Gamma_{\alpha,\beta} \ra = \sum_{\{s\}} (v_\alpha^T X A^{s_1} \dots A^{s_{L_b}} v_\beta - v_\alpha^T A^{s_1} \dots A^{s_{L_b}} X v_\beta) |s_1 \dots s_{L_b} \ra = -\frac{1}{\sqrt{2}} [(-1)^\alpha - (-1)^\beta] |\Gamma_{\alpha,\beta} \ra ~,
\end{equation}
where in the last equality we used $X = X^T$ and $X v_\alpha = -(-1)^\alpha \frac{1}{\sqrt{2}} v_\alpha$.
Thus, $|\Gamma_{1,1} \ra$ and $|\Gamma_{2,2} \ra$ are eigenstates with energy $0$, while $|\Gamma_{1,2} \ra$ and $|\Gamma_{2,1} \ra$ have energy $\sqrt{2}$ and $-\sqrt{2}$ respectively.

It is interesting to note that $\Gamma_{\alpha,\beta}$ are also eigenstates of $H'$ defined in Eq.~(\ref{eq:Hprime}).
From the diagonalization of $H'$ in Eq.~(\ref{eq:Hprime+-0}), any eigenstate of $H'$ must have energy which is an integer multiple of $\sqrt{2}$.
However, the fact that $H |\Gamma_{\alpha,\beta} \ra = H' |\Gamma_{\alpha,\beta} \ra$ only guarantees that $H' |\Gamma_{\alpha,\beta} \ra$ satisfy the Rydberg constraints, and we needed additional arguments to show that $|\Gamma_{\alpha,\beta} \ra$ are eigenstates of $H'$.

As an alternative proof, we can also write the OBC states $|\Gamma_{\alpha,\beta} \ra$ in the $\bm{\pm}, \bm{0}$ basis introduced in Eq.~(\ref{eq:+-0basis}).
Convenient MPS matrices in this basis are given in Eq.~(\ref{eq:Amatr+-0}), and the corresponding termination vectors (obtained using the gauge transformation that produced the convenient matrices) are $\tilde{v}_1 = V v_1 = (\sqrt{2}, 0)^T$ and $\tilde{v}_2 = V v_2 = (0, \sqrt{2})^T$.
It is now easy to see that for a product basis vector to be present in the expansion of $|\Gamma_{\alpha,\beta} \ra$, the leftmost non-$\bm{0}$ block-site must have $\bm{+}$ if $\alpha=1$ and $\bm{-}$ if $\alpha=2$, while the rightmost non-$\bm{0}$ site must have $\bm{-}$ if $\beta=1$ and $\bm{+}$ if $\beta=2$.
Similar to the case in PBC, properties $(A^{\bm{+}})^2 = (A^{\bm{-}})^2 = 0$ and $A^{\bm{0}} = 1$ imply that the $\bm{+}$'s and $\bm{-}$'s must alternate while allowing intervening $\bm{0}$'s.
Hence, we conclude that $N_{\bm{+}} - N_{\bm{-}} = 0$ for the $|\Gamma_{\alpha, \alpha} \ra$ states, while $N_{\bm{+}} - N_{\bm{-}} = 1$ or $-1$ for the $|\Gamma_{1,2} \ra$
or $|\Gamma_{2,1} \ra$ respectively, which reproduces the eigenvalues under $H'$ obtained earlier.

\section{Symmetries of the exact scar states}\label{sec:symmetries}

Here we derive the symmetry properties of the exact scar states listed in the main text.
We start with the exact scar states $|\Phi_i \ra$ in PBC.
For $L$ even, the inversion $I$ defined in the main text, $I: j \to L - j + 1$, is relative to a bond center and is not broken.
In the MPS representation, we have 
\begin{equation}
I |\Phi_1 \ra = \sum_{\{\sigma\}} \tr[B_I^{\sigma_1} C_I^{\sigma_2} \dots B_I^{\sigma_{L-1}} C_I^{\sigma_L}] |\sigma_1 \dots \sigma_L \ra ~,
\label{eq:IPhi1}
\end{equation}
where $B_I^\sigma \equiv [C^\sigma]^T$ and $C_I^\sigma \equiv [B^\sigma]^T$.
Consider now a $2 \times 2$ matrix $X_I \equiv i \sigma_y$ and a $3 \times 3$ matrix $Y_I \equiv \text{diag}(-1, -1, 1)$.
These satisfy $X_I B_I^\sigma Y_I^{-1} = B^\sigma$ and $Y_I C_I^\sigma X_I^{-1} = -C^\sigma$ and give us an MPS gauge transformation that proves $I |\Phi_1 \ra = (-1)^{L_b} |\Phi_1 \ra$.
For $|\Phi_2 \ra \equiv T_x |\Phi_1 \ra$, note that since $I T_x = T_x^{-1} I$ and $T_x^2|\Phi_i\ra=|\Phi_i\ra$, we have $I |\Phi_2 \ra  = (-1)^{L_b} |\Phi_2 \ra$.

While $\Cph$ is not a symmetry of $H$, our states are in fact eigenstates of $\Cph$.
Indeed, in terms of MPS, 
\begin{equation}
\Cph |\Phi_1 \ra = \sum_{\{\sigma\}} \tr[B_c^{\sigma_1} C_c^{\sigma_2} \dots B_c^{\sigma_{L-1}} C_c^{\sigma_L}] |\sigma_1 \dots \sigma_L \ra~,
\end{equation}
where $B_c^0 = -B^0$, $B_c^1 = B^1$, $C_c^0 = -C^0$, and $C_c^1 = C^1$.
Consider a $2 \times 2$ matrix $X_c \equiv \sigma_z$ and a $3 \times 3$ matrix $Y_c \equiv \text{diag}(-1, 1, -1)$.
Then applying the gauge transformation $X_c B_c^\sigma Y_c^{-1} = B^\sigma$, $Y_c C_c^\sigma X_c^{-1} = -C^\sigma$ proves $\Cph |\Phi_1 \ra = (-1)^{L_b} |\Phi_1 \ra$.
For $|\Phi_2 \ra$, noting that $\Cph T_x = T_x \Cph$, we conclude that $\Cph |\Phi_2 \ra  = (-1)^{L_b} |\Phi_2 \ra$.

Next we derive the symmetry properties of the exact scar states in OBC.
Under the inversion, 
\begin{equation}
    I |\Gamma_{\alpha,\beta} \ra = \sum_{\{\sigma\}} v_\beta^T B_I^{\sigma_1} C_I^{\sigma_2} \dots B_I^{\sigma_{L-1}} C_I^{\sigma_L} v_\alpha |\sigma_1 \dots \sigma_L \ra ~.
\end{equation}
Recall that $v_1 = (1, 1)^T$ and $v_2 = (1, -1)^T$.
We can perform the same gauge transformation as in PBC using matrices $X_I$ and $Y_I$.
The boundary vectors transform as $X_I v_1 = v_2$ and $X_I v_2 = -v_1$, and using also $X_I^{-1} = X_I^T$, we conclude that $I |\Gamma_{1,2} \ra = (-1)^{L_b-1} |\Gamma_{1,2} \ra$ and $I |\Gamma_{2,1} \ra = (-1)^{L_b-1} |\Gamma_{2,1} \ra$; while $I| \Gamma_{1,1} \ra = (-1)^{L_b} |\Gamma_{2,2} \ra$ and $I |\Gamma_{2,2} \ra = (-1)^{L_b} |\Gamma_{1,1} \ra$.

We also obtain that $\Cph |\Gamma_{1,2} \ra = (-1)^{L_b} |\Gamma_{2,1} \ra$ and $\Cph |\Gamma_{1,1} \ra = (-1)^{L_b} |\Gamma_{2,2} \ra$, etc, by employing the same gauge transformation as in PBC with matrices $X_c$ and $Y_c$ and noting that $X_c v_1 = v_2$ and $X_c v_2 = v_1$.

\section{Calculations of norms and overlaps of $|\Phi_1\ra$ and $|\Phi_2\ra$}
Calculations with the MPS state $|\Phi_1 \ra$ simplify in the blocked representation introduced in Sec.~\ref{app:proofPBC}.
They heavily use the associated transfer matrix defined as
\begin{align}
E_A = \sum_s (A^s)^* \otimes A^s = 
\begin{pmatrix}
2 & 0 & 0 & 1 \\
0 & 0 & -1 & 0 \\
0 & -1 & 0 & 0 \\
1 & 0 & 0 & 2
\end{pmatrix} ~,
\label{eqn:EA}
\end{align}
and we immediately get
\begin{align}
E_A^m = \frac{1}{2} 
\begin{pmatrix}
1 + 3^m & 0 & 0 & -1 + 3^m \\
0 & 1 + (-1)^m & -1 + (-1)^m & 0 \\
0 & -1 + (-1)^m & 1 + (-1)^m & 0 \\
-1 + 3^m & 0 & 0 & 1 + 3^m
\end{pmatrix} ~.
\end{align}
The norm of the state is $\la \Phi_1 | \Phi_1 \ra = \tr[E_A^{L_b}] = 3^{L_b} + 2 + (-1)^{L_b}$.
Since $|\Phi_2 \ra = T_x |\Phi_1 \ra$, we have $\la \Phi_2 | \Phi_2 \ra = \la \Phi_1 | \Phi_1 \ra$.

To calculate the overlap between $|\Phi_1 \ra$ and $|\Phi_2 \ra$ used in the main text, it is convenient to introduce blocked matrices $D^{(\sigma_1 \sigma_2)} = C^{\sigma_1} B^{\sigma_2}$. 
Specifically,
\begin{align}
D^O = \begin{pmatrix}
0 & -1 & 0 \\
1 & 0 & 0 \\
0 & 0 & 0
\end{pmatrix} ~, \quad
D^R = \begin{pmatrix}
-\sqrt{2} & 0 & -\sqrt{2} \\
0 & 0 & 0 \\
0 & 0 & 0
\end{pmatrix} ~, \quad
D^L = \begin{pmatrix}
\sqrt{2} & 0 & 0\\
0 & 0 & 0\\
-\sqrt{2} & 0 & 0
\end{pmatrix} ~.
\label{eq:Dmatr}
\end{align}
To calculate $\la \Phi_1 | \Phi_2 \ra$, we need the transfer matrix
\begin{align}
E_{AD} = \sum_s (A^s)^* \otimes D^s = 
\begin{pmatrix}
\sqrt{2}D^{R} & -D^{O}  \\
D^{O} & -\sqrt{2}D^{L} \\
\end{pmatrix} ~,
\label{eqn:EAD}
\end{align}
which is a $6 \times 6$ matrix with eigenvalues $-\sqrt{2}\!-\!1$, $-\sqrt{2}\!-\!1$, $0$, $0$, $\sqrt{2}\!-\!1$, and $\sqrt{2}\!-\!1$.
We therefore obtain $\la \Phi_1 | \Phi_2 \ra = \tr[E_{AD}^{L_b}]=2[(\sqrt{2}-1)^{L_b}+(-1)^{L_b}(\sqrt{2}+1)^{L_b}]$ quoted in the main text.

\section{One-site and two-site reduced density matrices in PBC for finite $L_b$}
Obtaining the one-site reduced density matrix of the exact states in PBC is a simple exercise in MPS calculations.
For concreteness, let us consider $|\Phi_1 \ra$. 
We define generalized transfer matrices $E_B^{\sigma \sigma^\prime} \equiv (B^{\sigma})^* \otimes B^{\sigma^\prime}$ and $E_C^{\sigma \sigma^\prime} \equiv (C^{\sigma})^* \otimes C^{\sigma^\prime}$.
The ordinary transfer matrices $E_B$ and $E_C$ defined earlier are related to these as $E_B=\sum_{\sigma}E_B^{\sigma\sigma}$ and $E_C=\sum_{\sigma}E_C^{\sigma\sigma}$.
We can now obtain the matrix elements of the one-site density matrix on the odd sites as $\la \sigma^\prime | \rho_{[1]}^{\text{one-site}} | \sigma \ra = \tr[E_B^{\sigma \sigma^\prime} E_C E_A^{L_b-1}] / \tr[E_A^{L_b}]$.
We find
\begin{equation}
\rho_{[1]}^{\text{one-site}} = \frac{2 \cdot 3^{L_b-1} + 1 + (-1)^{L_b}}{Z} |0 \ra \la 0| + \frac{3^{L_b-1} + 1}{Z} |1 \ra \la 1| ~,\quad\quad\quad Z = 3^{L_b} + 2 + (-1)^{L_b}~.
\end{equation}
On the even sites, the matrix elements are given as $\la \sigma^\prime | \rho_{[2]}^{\text{one-site}} | \sigma \ra = \tr[E_B E_C^{\sigma \sigma^\prime} E_A^{L_b-1}] / \tr[E_A^{L_b}]$.
It is easy to verify that we indeed have $\rho_{[1]}^{\text{one-site}} = \rho_{[2]}^{\text{one-site}}$. 
For even $L_b$, the one-site density matrix is $\rho_{[1]}^{\text{one-site}} = \rho_{[2]}^{\text{one-site}} = \frac{2}{3} |0 \ra \la  0| + \frac{1}{3} |1 \ra \la 1|$; while for odd $L_b$, it is essentially the same but with an exponentially small correction.

For the two-site reduced density matrix on sites $1$ and $2$, the matrix elements are given as $\la \sigma_1^\prime \sigma_2^\prime | \rho_{[1,2]} ^{\text{two-site}}| \sigma_1 \sigma_2 \ra 
= \tr[E_A^{(\sigma_1\sigma_2) (\sigma_1^\prime \sigma_2^\prime)} E_A^{L_b-1}] / \tr[E_A^{L_b}]$, where $E_A^{(\sigma_1\sigma_2)(\sigma_1^\prime \sigma_2^\prime)} = (A^{(\sigma_1 \sigma_2)})^*\otimes A^{(\sigma_1^\prime \sigma_2^\prime)}$, giving us
\begin{align}
\rho_{[1,2]}^{\text{two-site}} = & \frac{3^{L_b-1} + (-1)^{L_b}}{Z} | 00 \ra \la 00| + \frac{3^{L_b-1} + 1}{Z} (|01 \ra \la  01| + |10 \ra \la 10|) + \frac{-1 + (-1)^{L_b}}{3 Z} (|01 \ra \la 10| + |10 \ra \la 01|) ~.
\end{align}

On sites $2$ and $3$, the matrix elements of the two-site reduced density matrix are $\la \sigma_2^\prime \sigma_3^\prime |\rho_{[2,3]}^{\text{two-site}}|\sigma_2 \sigma_3 \ra = \tr[E_B E_D ^{(\sigma_2 \sigma_3) (\sigma_2^\prime \sigma_3^\prime)}E_C E_A^{L_b-2}]/\tr[E_A^{L_b}]$, where $E_D ^{(\sigma_2 \sigma_3)(\sigma_2^\prime \sigma_3^\prime)}\equiv (C^{\sigma_2}B^{\sigma_3})^*\otimes (C^{\sigma_2^\prime}B^{\sigma_3^\prime})$.
We find
\begin{align}
    \rho_{[2,3]}^{\text{two-site}}=\frac{3^{L_b-1}+(-1)^{L_b}}{Z}|00\ra\la00|+\frac{3^{L_b-1}+1}{Z}(|01\ra\la 01|+|10\ra\la 10|) 
    +\frac{1-3^{L_b-2}}{Z}(|01\ra\la 10|+|10\ra\la 01|)~.
\end{align}
For large $L_b$, these reduce to expressions in the main text.

As discussed in the main text, one-site observables cannot detect translation symmetry breaking, while the two-site observable $|0_j1_{j+1}\ra\la 1_j0_{j+1}|+ \hc$ can detect the $T_x$ breaking.
Another common observable in experiment and numerical studies---``domain wall number'' $P_j P_{j+1}$---has expectation value $1/3$ for any $j$ (and $L \rightarrow \infty$) and hence does not detect the translation symmetry breaking.
Interestingly, the Gibbs ensemble in the thermodynamic limit gives $\la P_j P_{j+1} \ra_{T=\infty} = \phi/(\phi + 2) \approx 0.4472$, which again directly shows the non-ETH behavior of $|\Phi_1 \ra$.

\section{Formulas for local energies in the exact scar states in OBC}
It is an easy exercise in MPS calculations to obtain expectation values of the local energy $\la X_j \ra_{\alpha,\beta} \equiv \la \Gamma_{\alpha,\beta} | X_j | \Gamma_{\alpha,\beta} \ra / \la \Gamma_{\alpha,\beta} | \Gamma_{\alpha,\beta} \ra$ in the OBC exact eigenstates defined in the main text.
The essential ingredients are generalized transfer matrices $E_{XB} = B^0 \otimes B^1 + B^1 \otimes B^0$ and $E_{XC} = C^0 \otimes C^1 + C^1 \otimes C^0$, as well as ordinary transfer matrices $E_B = B^0 \otimes B^0 + B^1 \otimes B^1$ and $E_C = C^0 \otimes C^0 + C^1 \otimes C^1$, where we have already used the fact that all our matrices $B^\sigma, C^\sigma$ are real.
Note that $E_B E_C = E_A$, which is the transfer matrix used in the blocked formulation and given in Eq.~(\ref{eqn:EA}).
We also define boundary vectors $e_\alpha = v_\alpha \otimes v_\alpha$, where $\alpha = 1, 2$ (here also using that our vectors $v_\alpha$ are real).
Parameterizing our terminations as $v_\alpha = (1, (-1)^{\alpha-1})$, $\alpha = 1, 2$, we obtain the norms as
\begin{align}
\la \Gamma_{\alpha,\beta} | \Gamma_{\alpha,\beta} \ra = e_\alpha^T E_A^{L_b} e_\beta = 2 \Big[(-1)^{L_b+\alpha+\beta} + 3^{L_b} \Big] ~.
\end{align}

For the energy calculations, at site $j = 2b-1$, $b = 1 \dots L_b$, we have 
$\la \Gamma_{\alpha,\beta} | X_j | \Gamma_{\alpha,\beta} \ra = 
e_\alpha^T E_A^{b-1} E_{XB} E_C E_A^{L_b-b} e_\beta$; while at site $j = 2b$, we have 
$\la \Gamma_{\alpha,\beta} | X_j | \Gamma_{\alpha,\beta} \ra = 
e_\alpha^T E_A^{b-1} E_B E_{XC} E_A^{L_b-b} e_\beta$.
We obtain
\begin{align}
\la X_{2b-1} \ra_{\alpha,\beta} = \la X_{2b} \ra_{\alpha,\beta} = \frac{\sqrt{2}}{1 + (-1)^{L_b + \alpha + \beta} 3^{-L_b}} \Big[ (-1)^\alpha (-1)^b 3^{-b} + (-1)^\beta (-1)^{L_b - b} 3^{-L_b + b - 1}  \Big] ~.
\end{align}
These are plotted in the main text. 
Interestingly, we can relate these states that differ by their terminations only at one edge by an action of a local two-site operator near that edge. 
For example, 
\begin{align}
|\Gamma_{1,2} \ra = \mathbf{1}_{[1, \dots, L-2]} \otimes \Big(&
\frac{1}{\sqrt{2}} |00 \ra \la 01| - \frac{1}{\sqrt{2}} |00 \ra \la 10|  + |01 \ra \la 01| - |10 \ra \la 10| \Big)_{[L-1,L]} |\Gamma_{1,1} \ra ~. \nonumber
\end{align}
(However, note that the operator achieving this is not unique.)

It is easy to check that the expectation value of the total energy $\la H \ra_{\alpha,\beta} = \sum_{j=1}^L \la X_j \ra_{\alpha,\beta}$ is $0$ if $\alpha = \beta$, while $\la H \ra_{1,2} = \sqrt{2} $ and $\la H \ra_{2,1} = -\sqrt{2} $.
In fact, the states are exact eigenstates and these expectation values are the corresponding eigenvalues, as we showed in Sec.~\ref{app:proofOBC}.

\section{Entanglement spectra of exact eigenstates in OBC}
To obtain the entanglement spectrum for the states $|\Gamma_{\alpha,\beta} \ra$, we follow the procedure in Refs.~\cite{ciracEntanglement2011,moudgalyaExact2018,moudgalyaEntanglement2018}.
First, we consider the entanglement cut between the sites $2b$ and $2b+1$.
We form $2 \times 2$ Gram matrix $\mathcal{L}_{2 \times 2}$ reshaped from $e_\alpha^T E_A^b$ and $2 \times 2$ Gram matrix $\mathcal{R}_{2 \times 2}$ reshaped from $E_A^{L_b-b} e_\beta$, and then obtain an effective matrix
\begin{equation}
\mathcal{S}_{2 \times 2} = \frac{\mathcal{L}_{2 \times 2} \mathcal{R}_{2 \times 2}}{\la \Gamma_{\alpha,\beta} | \Gamma_{\alpha,\beta} \ra} = \frac{1}{2} \begin{pmatrix}
1 & \frac{(-1)^{\alpha+1} 3^{L_b-b} + (-1)^{\beta+1} 3^b}{(-1)^{L_b+\alpha+\beta} + 3^{L_b}} \\
\frac{(-1)^{\alpha+1} 3^{L_b-b} + (-1)^{\beta+1} 3^b}{(-1)^{L_b+\alpha+\beta} + 3^{L_b}} & 1
    \end{pmatrix}
\end{equation}
whose eigenvalues are the same as eigenvalues of the reduced density matrix.
We therefore obtain the entanglement spectrum as
\begin{equation}
s_{1,2} = \frac{1}{2} \left( 1 \pm \frac{(-1)^{\alpha+1} 3^{L_b-b} + (-1)^{\beta+1} 3^b}{(-1)^{L_b+\alpha+\beta} + 3^{L_b}} \right) ~.
\end{equation}
For large subsystem size $b$ and in the thermodynamic limit, e.g., where we take $L_b \to \infty$, $b \to \infty$, while fixing the ratio $b/L_b = f < 1$, or where we take $L_b \to \infty$ first and then $b \to \infty$, the entanglement spectrum approaches $s_{1,2} \to 1/2$ independent of the terminations.

For the entanglement cut between sites $2b + 1$ and $2b + 2$, we need $3 \times 3$ Gram matrix $\mathcal{L}_{3 \times 3}$ reshaped from $e_\alpha^T E_A^b E_B$ and $3 \times 3$ Gram matrix $\mathcal{R}_{3 \times 3}$ reshaped from $E_C E_A^{L_b-b-1} e_\beta$.
The effective matrix that reproduces the entanglement spectrum is 
\begin{align}
    \mathcal{S}_{3 \times 3}=&\frac{\mathcal{L}_{3 \times 3}\mathcal{R}_{3 \times 3}}{\la \Gamma_{\alpha,\beta}|\Gamma_{\alpha,\beta} \ra}=  \frac{1}{3^{L_b}+(-1)^{L_b+\alpha+\beta}}\notag \\
    \times & \begin{pmatrix}
    \frac{5}{6}\cdot3^{L_b}+\frac{1}{2}\cdot(-1)^{L_b+\alpha+\beta} &
    \frac{1}{6}\cdot(-1)^{1+b}[(-1)^{\alpha}\cdot 3^{L_b-b}+9\cdot(-1)^{L_b+\beta}\cdot 3^{b}] & 
    -3^{L_b-1} \\[0.4cm]
    \frac{1}{2}\cdot(-1)^{1+b}[(-1)^{\alpha}\cdot 3^{L_b-b}+(-1)^{L_b+\beta}\cdot 3^{b}] &
     \frac{1}{6}\cdot 3^{L_b}+\frac{1}{2}\cdot (-1)^{L_b+\alpha+\beta} & 
     (-1)^{b+\alpha}\cdot 3^{L_b-b-1}\\[0.4cm]
    3^{L_b-1}
    & (-1)^{1+\beta+L_b-b}\cdot 3^{b} & 0
    \end{pmatrix}~.
\end{align}
The entanglement spectrum at any finite $L_b$, $b$ can be obtained from the eigenvalues of the above matrix. 
For large $b$ and in the thermodynamic limit, we have
\begin{eqnarray}
\mathcal{S} \to \begin{pmatrix}
\frac{5}{6} & 0 & -\frac{1}{3} \\
0 & \frac{1}{6} & 0 \\
\frac{1}{3} & 0 & 0
\end{pmatrix} ~,
\end{eqnarray}
which gives the entanglement spectrum $2/3$, $1/6$ and $1/6$ quoted in the main text.

\section{Comparison with the forward scattering approximation}
Here we follow Refs.~\cite{turnerWeak2018, turnerQuantum2018} to construct the foward scattering approximation (FSA) and compare with our alternative picture.
Our main goal is to compare performance of our multi-mode approximations (MMA) on top of the exact $E=0$ states and the FSA, in order to argue that our exact states and approximate quasiparticle constructions on top of these are relevant for the $Z_2$ scar states.
In the FSA, one constructs a ``variational" subspace, where one starts from $|Z_2 \ra \equiv |10 \dots 10 \ra$ and operates with $H^+ \equiv \sum_{j \in \text{even}} P_{j-1} \sigma^+_j P_{j+1} + \sum_{j \in \text{odd}} P_{j-1} \sigma^-_j P_{j+1}$ to form basis vectors $|n \ra = (H^+)^n |Z_2 \ra / \|(H^+)^n | Z_2 \ra\|$ for $n = 0, 1, \dots, L$.
It is easy to see that $|L \ra = |Z_2 ^\prime \ra \equiv |01 \dots 01 \ra$.  
One then projects the full Hamiltonian into this subspace and obtains an effective Hamiltonain $H_\text{FSA}$, which is an $(L+1) \times (L+1)$ matrix with basis $|n \ra$.
By construction, $H_\text{FSA}$ is bidiagonal. 
Diagonalzing $H_\text{FSA} = S E_\text{FSA} S^\dagger$, one obtains ``variational" energies $E_{\text{FSA}, i}$ and approximate wavefunctions $|\text{FSA}_i \ra = \sum_{n=0}^L S_{ni} | n \rangle$, where $i = 0, 1, \dots, L$. 

\begin{figure}
    \includegraphics[width=0.5\columnwidth]{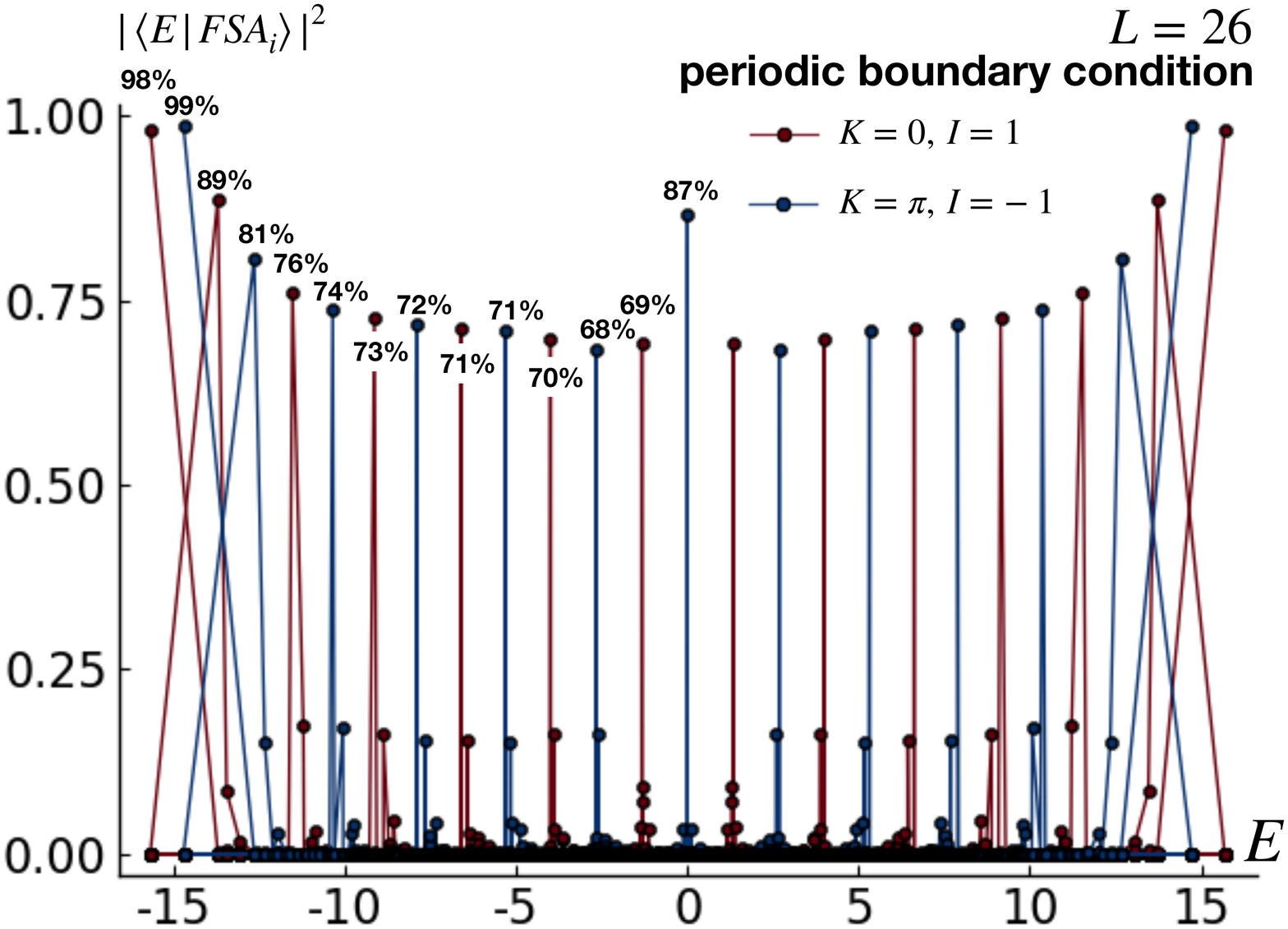}
    \caption{Overlaps between the FSA wavefunctions, $|\text{FSA}_i \ra$, $i = 0, 1, \dots, L$, and the eigenstates in the PBC chain with $L = 26$.
    The red lines are in the $K = 0$, $I = 1$ sector, while the blue lines are in the $K = \pi$, $I = -1$ sector.
    $|\text{FSA}_i \ra$ has the largest overlap with the $i$-th $Z_2$ scar state, and this value is listed for easy reference for $i = 0, 1, \dots, L/2$.} 
    
    \label{fig:FSA}
\end{figure}

We show the overlaps between the FSA wavefunctions and the eigenstates, $|\la E|\text{FSA}_i \ra|^2$, for $i = 0, 1, \dots, L$ in Fig.~\ref{fig:FSA}.
We also quote the overlap values on the ``matching'' primary $Z_2$ scar states, i.e., between $|\text{FSA}_i \ra$ and the $i$-th primary scar state, with both sets of states assumed ordered by energy.
In general, the FSA provides very good approximations for the primary $Z_2$ scar states, and, in particular, an extremely good approximation for the ground state and the scar state closest to the ground state.

It is interesting to note that the FSA variational space and the FSA states $|\text{FSA}_i \ra$ generated by the above procedure do not respect the translation and inversion symmetries of the Hamiltonian.
Instead, they mix the $K = 0$, $I = 1$ and $K = \pi$, $I = -1$ sectors.
However, each individual $|\text{FSA}_i \ra$ state generally has a very high weight on a particular symmetry sector.
In principle, one can fix this completely by including the symmetry-related counterparts in the variational basis, but we have not done such an embellishment and only followed the original procedure in Refs.~\cite{turnerWeak2018, turnerQuantum2018}.
On the other hand, our trial SMA and MMA wavefunctions are constructed with definite symmetry quantum numbers from the outset.

\section{Additional details and results of single-mode approximation and multi-mode approximation}

\subsection{Details of SMA for the $E \approx \pm 1.33$ scar states with $I = (-1)^{L_b-1}$ in the main text}
In the main text, we presented the ``single-mode approximation" (SMA) with the translation quantum number $(-1)^{L_b-1}$ and inversion quantum number $(-1)^{L_b-1}$ to capture the scar states with energy $E \approx \pm 1.33$.
For ease of reference, we remind the construction and then explain more details behind it: $|\Xi_1 \ra  = (|M_1 \ra - (-1)^{L_b} T_x |M_1 \ra)/\xi_1$, where $\xi_1$ provides normalization $\la \Xi_1|\Xi_1 \ra = 1$ and
\begin{equation}\label{eqn:M1}
    |M_1 \ra = \sum_{b=1}^{L_b} \tr[B^{\sigma_1} C^{\sigma_2} \dots M^{\sigma_{2b-1} \sigma_{2b}} \dots B^{\sigma_{L-1}} C^{\sigma_L}] |\sigma_1 \dots \sigma_L \ra ~, 
\end{equation}
with
\begin{align*}
M^{00} = \begin{pmatrix}
1 & 0 \\
0 & 1
\end{pmatrix} ~, \quad
M^{01} = \begin{pmatrix}
\mu_1 & 0 \\
\mu_2 & 0
\end{pmatrix} ~, \quad
M^{10} = \begin{pmatrix}
0 & 0 \\
-\mu_2 & \mu_1
\end{pmatrix} ~, \quad
M^{11} = 0_{2 \times 2} ~.
\end{align*}

We have chosen the ``excitation'' matrices $M$ to satisfy $M^{01} B^1 = 0_{2 \times 3}$ and $C^1 M^{10} = 0_{3 \times 2}$ so that the wavefunction automatically satisfies the Rydberg blockade constraint.
Furthermore, we have required that the matrices give the inversion quantum number opposite to the exact $E=0$ eigenstate $|\Phi_1 \ra$:
By examining the action of $I$ on $|M_1 \ra$ in the MPS language similar to Eq.~(\ref{eq:IPhi1}) and utilizing the same gauge transformation used in the discussion after Eq.~(\ref{eq:IPhi1}), we see that the desired inversion quantum number is achieved by requiring $X_I M_I^s X_I^{-1} = M^s$, where $M_I^{00} \equiv (M^{00})^T, M_I^{01} \equiv (M^{10})^T, M_I^{10} \equiv (M^{01})^T$.
Satisfying these conditions leads to the ansatz with two parameters $\mu_1$ and $\mu_2$ shown above.
In principle, the SMA wavefunction has a ``gauge redundancy'', i.e., property that $M^{\sigma_1 \sigma_2} \to M^{\sigma_1 \sigma_2} + [W, B^{\sigma_1} C^{\sigma_2}]$ does not change $|M_1 \ra$ for arbitrary $2 \times 2$ matrix $W$. 
We need to consider this redundancy to find the set of truly independent parameters.
In the present case, it happens that the above ansatz for the excitation matrices has independent parameters already.
Hence, the gauge redundancy does not reduce the number of the independent parameters. 
The optimal parameters are obtained by minimizing the energy fluctuation $\sigma^2_H \equiv \la \Xi_1|H^2|\Xi_1 \ra - \la \Xi_1|H|\Xi_1 \ra^2$.
The resulting optimal state is presented in the main text and also reproduced in Fig.~\ref{fig:qpall}.

We also note that by choosing $\mu_1^\prime = -\mu_1$ and $\mu_2^\prime = \mu_2$, we can obtain the opposite-energy counterpart, $|\Xi_1^\prime \ra \sim \Cph |\Xi_1 \ra$.
This can be seen using the gauge transformation introduced in the discussion of the action of $\Cph$ on the exact eigenstate $|\Phi_1 \ra$ in Sec.~\ref{sec:symmetries} and noting that the corresponding excitation matrices satisfy $(M^\prime)^s = X_c M_c^s X_c^{-1}$, where $M_c^{00} \equiv M^{00}, M_c^{01} \equiv -M^{01}, M_c^{10} \equiv -M^{10}$.

Finally, we can provide some intuition for the energetics of the SMA ansatz by working in the blocked language used in Sec.~\ref{app:proofPBC} utilizing the $\bm{\pm}, \bm{0}$ basis introduced in Eq.~(\ref{eq:+-0basis}).
Recall that in this basis $|\Phi_1 \ra$ is conveniently written using MPS matrices in Eq.~(\ref{eq:Amatr+-0}).
We can easily obtain excitation matrices in this representation by following the same steps that produced Eq.~(\ref{eq:Amatr+-0}); we find
\begin{align}
M^{\bm{+}} = \frac{\mu_1 + \sqrt{2}}{2} \begin{pmatrix}
1 & 0 \\
0 & 1
\end{pmatrix} ~, \quad
M^{\bm{-}} = \frac{\mu_1 - \sqrt{2}}{2} \begin{pmatrix}
1 & 0 \\
0 & 1
\end{pmatrix} ~, \quad
M^{\bm{0}} = \frac{1}{\sqrt{2}} \begin{pmatrix}
\mu_2 & \mu_1 + \mu_2 \\
\mu_1 - \mu_2 & -\mu_2
\end{pmatrix} ~.
\label{eq:Mmatr+-0}
\end{align}
We can now examine a wavefunction obtained by placing such an excitation on one block-site $b$.
The wavefunction has a part with $s_b = \bm{+}$, originating from $M^{\bm{+}}$, which on the rest of the system is basically $\tr[A A \dots A]$ and hence contains only configurations with equal numbers of $\bm{+}$ and $\bm{-}$ block-sites.
Hence, this part is an eigenstate of $H'$ introduced in Sec.~\ref{app:proofPBC}, see Eqs.~(\ref{eq:Hprime}) and (\ref{eq:Hprime+-0}), with eigenvalue $\sqrt{2}$.
Similarly, a part of the wavefunction with $s_b = \bm{-}$, originating from $M^{\bm{-}}$, on the rest of the system contains only configurations with equal numbers of $\bm{+}$ and $\bm{-}$ block-sites; hence, this part is an eigenstate of $H'$ with eigenvalue $-\sqrt{2}$.
Note that the amplitudes of the two parts are proportional to $(\mu_1 + \sqrt{2})/2$ and $(\mu_1 - \sqrt{2})/2$ respectively, and for the optimal parameters $\mu_1 \approx -1.09$, $\mu_2 \approx -0.63$ used in the main text, the latter amplitude is significantly larger.

We now examine a part of the wavefunction with $s_b = \bm{0}$, originating from $M^{\bm{0}}$, which we further subdivide as follows.
The upper-left subpart $\mu_2/\sqrt{2}$ in $M^{\bm{0}}$ by itself requires that on the rest of the system we have equal numbers of $\bm{+}$'s and $\bm{-}$'s, and the first such site to the right of $b$ must be $\bm{+}$.
Similarly, the lower-right subpart $-\mu_2/\sqrt{2}$ in $M^{\bm{0}}$ by itself requires that on the rest of the system we have equal numbers of $\bm{+}$'s and $\bm{-}$'s, and the first such site to the right of $b$ must be $\bm{-}$.
Each of these cases gives an eigenstate of $H'$ with eigenvalue $0$.
Next, the lower-left subpart $(\mu_1 - \mu_2)/\sqrt{2}$ in $M^{\bm{0}}$ by itself requires configurations on the rest of the system to be of the form $\dots {\bm{+}} \dots {\bm{-}} \dots {\bm{+}} \dots$, where ``$\dots$'' can contain any number of $\bm{0}$'s.
Such configurations contain one more $\bm{+}$ compared to $\bm{-}$, which gives an eigenstate of $H'$ with eigenvalue $\sqrt{2}$.
Similarly, the upper-right subpart $(\mu_1 + \mu_2)/\sqrt{2}$ in $M^{\bm{0}}$ by itself requires configurations on the rest of the system to be of the form $\dots {\bm{-}} \dots {\bm{+}} \dots {\bm{-}} \dots$, i.e., contain one less $\bm{+}$ compared to $\bm{-}$, which gives an eigenstate of $H'$ with eigenvalue $-\sqrt{2}$.

We can now use the fact that the expectation value of $H$ in this wavefunction coincides with the expectation value of $H'$, since the genuinely two-body parts in the writing of $H$ in Sec.~\ref{app:proofPBC} connect to outside of the Rydberg-constrained Hilbert space.
(As a side remark, we can obtain the action of $H$ on this wavefunction by first acting with $H'$ and then projecting into the Rydberg Hilbert space; in particular, one can see that such an excitation wavefunction is no longer exact eigenstate of $H$.)
By examining contributions to the expectation value of $H'$ from the above parts of the wavefunction, we can roughly understand the value of the trial energy $\approx -1.31$ obtained using these excitation matrices in the main text. 
Also, we can see that changing the sign of $\mu_1$ gives a trial state with opposite energy, in agreement with the formal argument using $\Cph$ given earlier.

Note that in the main text we formed plane wave superpositions of such localized excitations, and in the analysis here we are not attempting a quantitative match with the numerical results.
Also note that while this analysis provides a rough intuition for the trial energies, in the main text we optimized the SMA ansatz by {\it minimizing the variance}, for which we have less intuition.
Nevertheless, the above arguments provide an approximate picture where adding an excitation is like acting with a ladder operator raising or lowering eigenvalues of $H'$ in Eq.~(\ref{eq:Hprime+-0}), and developing this picture more precisely may provide a better understanding of the scar states in the PXP model away from $E=0$.

\subsection{SMA for the $E \approx \pm 2.66$ scar states with $I = (-1)^{L_b}$}
While in the main text we showed the multi-mode approximation (MMA) to capture other $Z_2$ scar states, here we can also try to use the SMA but with the symmetry quantum numbers $T_x = (-1)^{L_b}$ and $I = (-1)^{L_b}$.
Specifically, we can write an SMA wavefunction with such quantum numbers as 
$|\tilde{\Xi}_1 \ra = (|\tilde{M}_1 \ra + (-1)^{L_b} T_x |\tilde{M}_1 \ra)/\tilde{\xi}_1$, where again $\tilde{\xi}_1$ is the normalization factor and $|\tilde{M}_1 \ra$ has the same form as in Eq.~(\ref{eqn:M1}) but with matrices
\begin{align*}
\tilde{M}^{00} = \begin{pmatrix}
\sqrt{2} & -\sqrt{2} \tilde{\mu}_1 \\
\sqrt{2} \tilde{\mu}_1 & -\sqrt{2}
\end{pmatrix} ~, \quad
\tilde{M}^{01} = \begin{pmatrix}
-\tilde{\mu}_1 & 0 \\
-1 & 0
\end{pmatrix} ~, \quad
\tilde{M}^{10} = \begin{pmatrix}
0 & 0 \\
-1 & \tilde{\mu}_1
\end{pmatrix} ~.
\end{align*}
Similarly to the construction of $|M_1 \ra$, we have chosen the matrices $\tilde{M}^s$ to satisfy $\tilde{M}^{01} B^1 = 0_{2 \times 3}$ and $C^1 \tilde{M}^{10} = 0_{3 \times 2}$ but with $X_I \tilde{M}_I^s X_I^{-1} = -\tilde{M}^s$ to give the same inversion quantum number as $|\Phi_i \ra$.
We have also used the SMA gauge redundancy, $\tilde{M}^{\sigma_1\sigma_2} \to \tilde{M}^{\sigma_1\sigma_2} + [W, B^{\sigma_1}C^{\sigma_2}]$, to identify the truly independent parameters.
Finally, the number of the independent parameters was reduced by one by requiring $|\tilde{M}_1 \ra$ to be orthogonal to $|\Phi_1 \ra$ in the thermodynamic limit (we did not need to do this for $|M_1 \ra$ since it has different inversion quantum number and is automatically orthogonal to $|\Phi_1 \ra$).

The optimal parameter $\tilde{\mu}_1$ is obtained by minimizing the energy fluctuation; using system of length $L=26$, we find $\tilde{\mu}_1 = 0.89285$, $\la \tilde{\Xi}_1|H|\tilde{\Xi}_1 \ra = -2.4572$ and $\la \tilde{\Xi}_1|H^2|\tilde{\Xi}_1 \ra - \la \tilde{\Xi}_1|H|\tilde{\Xi}_1 \ra^2 = 0.3219$.
To obtain the positive energy counterpart, we can choose $\tilde{\mu}_1^\prime = -\tilde{\mu}_1$, which gives $|\tilde{\Xi}_1^\prime \ra \sim \Cph |\tilde{\Xi}_1 \ra$
(the argument is essentially identical to that for $|\Xi_1^\prime \ra \sim \Cph |\Xi_1 \ra$ at the end of the previous subsection).
The overlap of $|\tilde{\Xi}_1 \ra$ with the eigenstates is plotted in Fig.~\ref{fig:qpall}. 
While this state still has majority of the weight on the primary scar state with $E \approx -2.66$, the overlap is significantly worse than the multi-particle ansatz $|\Xi_2 \ra$ presented in the main text.
The wavefunction $|\tilde{\Xi}_1 \ra$ can be loosely viewed as a bound state of two quasiparticles, while $|\Xi_2 \ra$ can be viewed as a scattering state of two quasiparticles.
These results suggest that the scar states are better understood as essentially free quasiparticle states rather than bound states of quasiparticles.

\subsection{``Bond-dimension-3" SMA}
\begin{figure}
    \includegraphics[width=0.5\columnwidth]{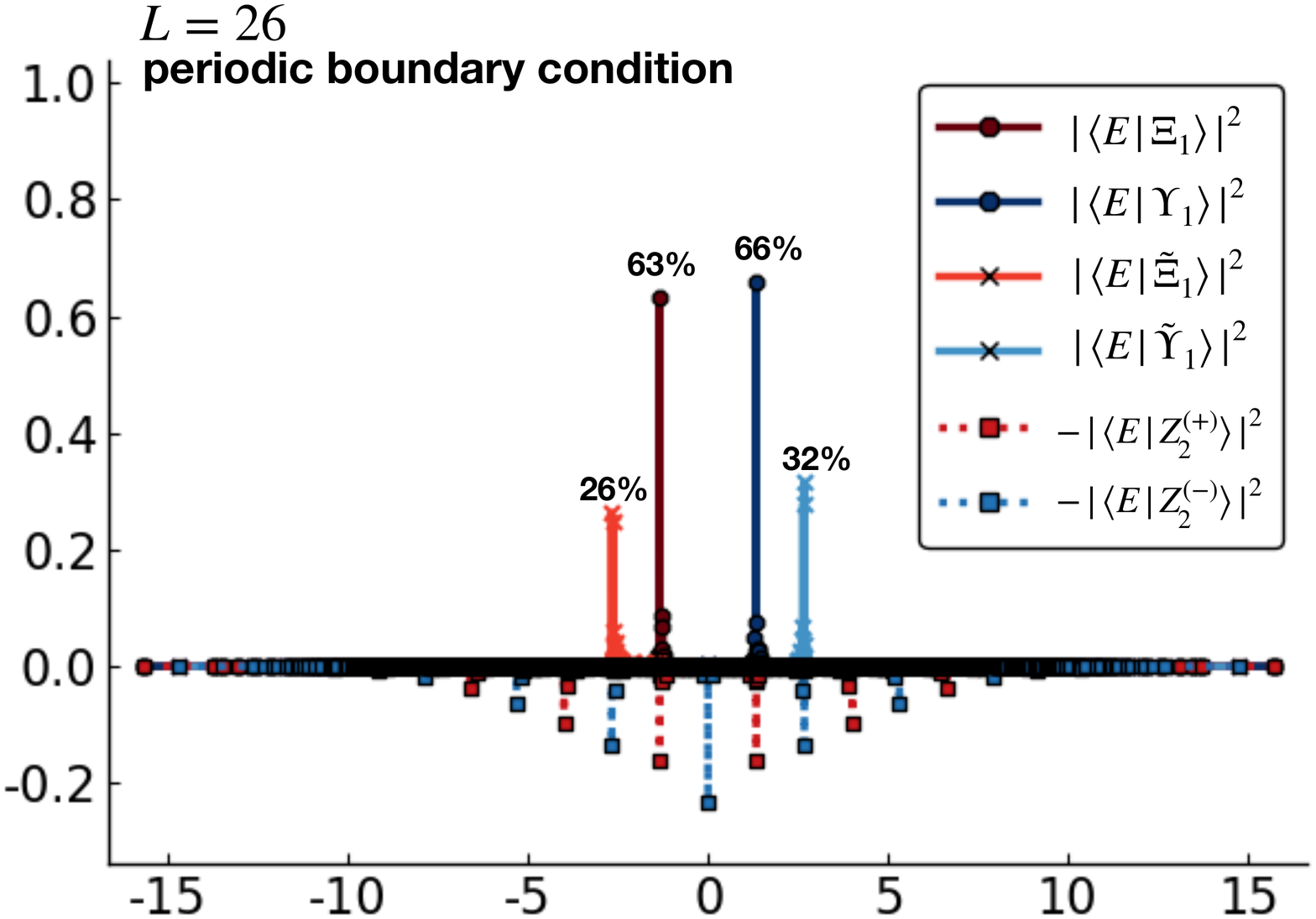}
    \caption{Overlaps of the SMA wavefunctions with eigenstates in the PBC chain of length $L=26$.
    Here the SMA wavefunctions are constructed using ``bond-dimension 2" ($|\Xi_1 \ra$ and $|\tilde{\Xi}_1 \ra$) and ``bond-dimension 3" ($|\Upsilon_1 \ra$ and $|\tilde{\Upsilon}_1 \ra$) ansatzes, with choices producing different symmetry sectors.
    The red lines and the non-tilde states label the $K=0$, $I=1$ sector; while the blue lines and the tilded states label the $K=\pi$, $I=-1$ sector.}
    \label{fig:qpall}
\end{figure}

The SMA wavefunctions $|\Xi_1 \ra$ and $|\tilde{\Xi}_1 \ra$ are constructed by ``exciting" the matrices $B$ and $C$ consecutively, where the ``quasiparicle excitation" matrix $M$ has bond-dimension 2.
We therefore call these wavefunctions ``bond-dimension-2" ansatzes. 
On the other hand, one can also think of excitations on consecutive matrices $C$ and $B$, which will give the quasiparticle excitation matrix with bond-dimension 3.
Such ansatzes will have more variational parameters and can potentially be better approximations.
The bond-dimension-3 SMA capturing the $E \approx \pm 1.33$ scar states with quantum numbers $T_x = (-1)^{L_b-1}$ and $I = (-1)^{L_b-1}$ is
$|\Upsilon_1 \ra = (|N_1 \ra - (-1)^{L_b} T_x |N_1 \ra)/\upsilon_1$, where $\upsilon_1$ is the normalization factor and 
\begin{equation}\label{eqn:N1}
    |N_1 \ra = \sum_{b=1}^{L_b} \tr[B^{\sigma_1} C^{\sigma_2} \dots B^{\sigma_{2b-1}} N^{\sigma_{2b} \sigma_{2b+1}} C^{\sigma_{2b+2}} \dots B^{\sigma_{L-1}} C^{\sigma_L}] |\sigma_1 \dots \sigma_L \ra ~, 
\end{equation}
with
\begin{align*}
N^{00} = \begin{pmatrix}
1 + \sqrt{2}\, \nu_7 & 0 & 2 \sqrt{2}\, \nu_6 \\
0 & 1 - \sqrt{2}\, \nu_7 & 2 \sqrt{2}\, \nu_5 \\
-2 \sqrt{2}\, \nu_6 & 2 \sqrt{2}\, \nu_5 & \nu_1
\end{pmatrix} ~,~~
N^{01} = \begin{pmatrix}
\nu_2 & 2\nu_6 & \nu_2 \\
\nu_6 + \nu_7 & \nu_3 & \nu_6 + \nu_7 \\
\nu_2 - \nu_5 & \nu_4 & \nu_2 - \nu_5
\end{pmatrix} ~,~~
N^{10} = \begin{pmatrix}
\nu_2 & \nu_6 + \nu_7 & -\nu_2 - \nu_5 \\
2\nu_6 & \nu_3 & -\nu_4 \\
-\nu_2 & -\nu_6 - \nu_7 & \nu_2 + \nu_5
\end{pmatrix} ~.
\end{align*}
We obtained these matrices by requiring $N^{01} C^1 = 0_{3 \times 2}$, $B^1 N^{10} = 0_{2 \times 3}$ to ensure the Rydberg constraints.
To obtain the desired inversion quantum number, we also required $Y_I N_I^s Y_I^{-1} = N^s$, where $N_I^{00} \equiv (N^{00})^T$, $N_I^{01} \equiv (N^{10})^T$, $N_I^{10} \equiv (N^{01})^T$, and $Y_I$ is the matrix for the gauge transformation used in our discussion of $I$ in Sec.~\ref{sec:symmetries}.
Finally, we also used the SMA gauge redundancy, $N^{\sigma_1\sigma_2} \to N^{\sigma_1\sigma_2} + [W,C^{\sigma_1}B^{\sigma_2}]$, to find truly independent parameters as shown above.
For system size $L = 26$, we find the optimal parameters $\nu_1 = 0.507183$, $\nu_2 = 0.60202$, $\nu_3 = 0.625366$, $\nu_4 = 0.264115$, $\nu_5 = -0.00128607$, $\nu_6 = 0.0228075$, and $\nu_7 = 0.42342$.
The trial energy is $\la \Upsilon_1|H|\Upsilon_1 \ra = 1.3396$ and the energy fluctuation is $\la \Upsilon_1|H^2|\Upsilon_1 \ra - \la \Upsilon_1|H|\Upsilon_1 \ra^2 = 0.007201$, which is a more accurate approximation than the $|\Xi_1 \ra$ SMA state in the main text.
We can see from Fig.~\ref{fig:qpall} that the overlap with the primary $Z_2$ scar state with energy $E \approx 1.33$ is $66\%$, which is higher than the bond-dimension-2 ansatz $|\Xi_1 \ra$.
To obtain the negative energy counterpart, one can simply change the signs of $\nu_{2,3,5}$ and obtain $|\Upsilon_1^\prime \ra \sim \Cph |\Upsilon_1 \ra$, which is deduced by applying the discussion of $\Cph$ in Sec.~\ref{sec:symmetries} and using $(N^\prime)^s = Y_c N_c^s Y_c^{-1}$, where $N_c^{00} \equiv N^{00}, N_c^{01} \equiv -N^{01}, N_c^{10} \equiv -N^{10}$.

Similarly, we can construct the bond-dimension-3 SMA in the symmetry sector $T_x = (-1)^{L_b}$ and $I = (-1)^{L_b}$ as $|\tilde{\Upsilon}_1 \ra = (|\tilde{N}_1 \ra + (-1)^{L_b} T_x |\tilde{N}_1 \ra)/\tilde{\upsilon}_1$, where $\tilde{\upsilon}_1$ is the normalization factor and $|\tilde{N}_1 \ra$ has the same form as in Eq.~(\ref{eqn:N1}) but with matrices
\begin{align*}
\tilde{N}^{00} = \begin{pmatrix}
0 & 2 \sqrt{2} & 2 \sqrt{2}\, \tilde{\nu}_4 \\
-2 \sqrt{2} & 0 & 2 \sqrt{2}\, \tilde{\nu}_3 \\
2 \sqrt{2}\; \tilde{\nu}_4 & -2 \sqrt{2}\, \tilde{\nu}_3 & 0
\end{pmatrix} ~,~~
\tilde{N}^{01} = \begin{pmatrix}
-1 & -2 \tilde{\nu}_4 & -1 \\
\tilde{\nu}_4 & \tilde{\nu}_1 & \tilde{\nu}_4 \\
-1 + \tilde{\nu}_3 & \tilde{\nu}_2 & -1 + \tilde{\nu}_3
\end{pmatrix} ~,~~
\tilde{N}^{10} = \begin{pmatrix}
1 & -\tilde{\nu}_4 & -1 - \tilde{\nu}_3 \\
2\tilde{\nu}_4 & -\tilde{\nu}_1 & \tilde{\nu}_2 \\
-1 & \tilde{\nu}_4 & 1 + \tilde{\nu}_3
\end{pmatrix} ~.
\end{align*}
In addition to satisfying the Rydberg constraint and giving the inversion quantum number $I = (-1)^{L_b}$, the matrices are chosen such that $|\tilde{\Upsilon}_1 \ra$ is orthogonal to $|\Phi_1 \ra$ in the thermodynamic limit.
Using system size $L=26$, we find the optimal parameters $\tilde{\nu}_1 = 2.59334$, $\tilde{\nu}_2 = 1.48065$, $\tilde{\nu}_3 = 0.0615383$, and $\tilde{\nu}_4 = -0.992914$, with the trial energy $\la \tilde{\Upsilon}_1 | H | \tilde{\Upsilon}_1 \ra = 2.5594$, and energy fluctuation $\la \tilde{\Upsilon}_1 | H^2 | \tilde{\Upsilon}_1 \ra - \la \tilde{\Upsilon}_1 | H | \tilde{\Upsilon}_1 \ra^2 = 0.18591$.
To obtain the corresponding negative-energy trial state $|\tilde{\Upsilon}^\prime_1 \ra \sim \Cph |\tilde{\Upsilon}_1 \ra$, one changes the signs of $\tilde{\nu}_2$ and $\tilde{\nu}_4$.
We again see from Fig.~\ref{fig:qpall} that the overlap with the $Z_2$ primary scar state at $E \approx 2.66$ is higher than the bond-dimension-2 ansatz $|\tilde{\Xi}_1 \ra$, but significantly worse than the MMA wavefunction $|\Xi_2 \ra$.

\begin{figure}
    \includegraphics[width=0.5\columnwidth]{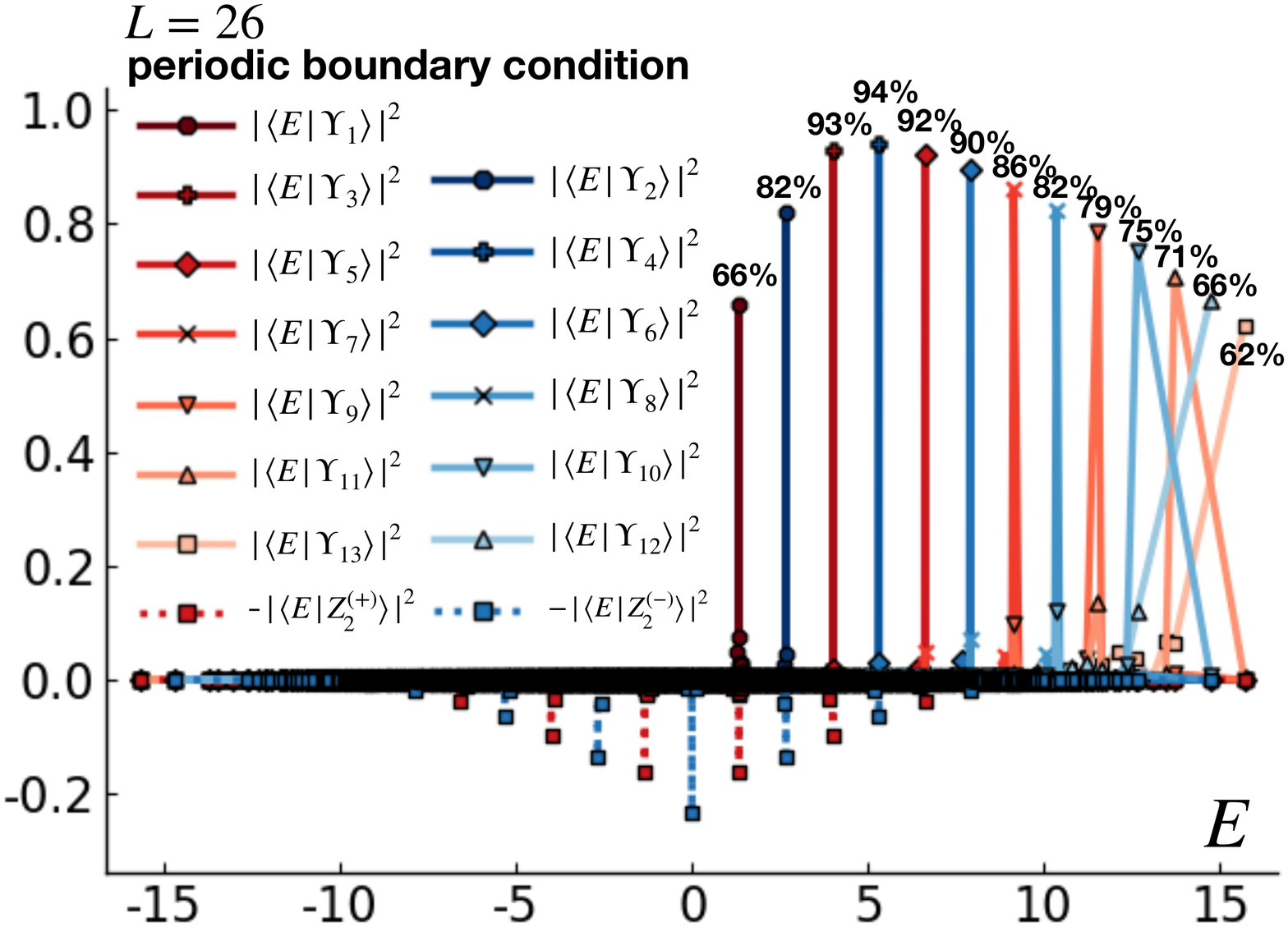}
    \caption{Overlaps between the MMA wavefunctions $|\Upsilon_n \ra$ with eigenstates in the PBC chain with $L=26$.
    The quasiparticle $N$ matrices are chosen from the optimal ``bond-dimension 3" SMA wavefunction.
    These wavefunctions represent the simplest scattering states of the quasiparticles with hard-core exclusions.
    }
    \label{fig:qpbd3}
\end{figure}

In fact, we can again construct the corresponding ``bond-dimension-3" MMA wavefunctions: $|\Upsilon_n \ra = (|N_n \ra + (-1)^{L_b + n} T_x |N_n \ra)/\upsilon_n$, where
\begin{align}
    |N_n \ra = & \sideset{}{'}\sum_{b_1, \dots, b_n = 1}^{L_b} \tr[B^{\sigma_1} C^{\sigma_2} \dots N^{\sigma_{2b_1} \sigma_{2b_1+1}} \dots 
     N^{\sigma_{2b_n} \sigma_{2b_n+1}} \dots B^{\sigma_{L-1}} C^{\sigma_L}] |\sigma_1 \dots \sigma_L \ra ~,
\end{align}
the summation is constrained to have all $b_i$ distinct, and $\upsilon_n$ is the normalization factor.
We take the $N$ matrices from the optimal result of $|\Upsilon_1 \ra$ and examine the overlaps of $|\Upsilon_n \ra$ with the eigenstates, in particular, with the $Z_2$ primary scar states in Fig.~\ref{fig:qpbd3}.
Similar to the bond-dimension-2 results presented in the main text, these bond-dimension-3 MMA wavefunctions have symmetry quantum numbers $T_x=I=(-1)^{L_b+n}$ and, remarkably, capture the primary scar states with even higher fidelity with more quasiparticles, up to $n \approx L_b/4$.
In this case, even the ground state and the primary scar states near the ground state are approximated fairly well compared to the results from the bond-dimension-2 ansatzes $|\Xi_n \ra$. 

\section{Size dependence of the bipartite entanglement entropy of SMA and MMA}
\begin{figure}
    \includegraphics[width=0.5\columnwidth]{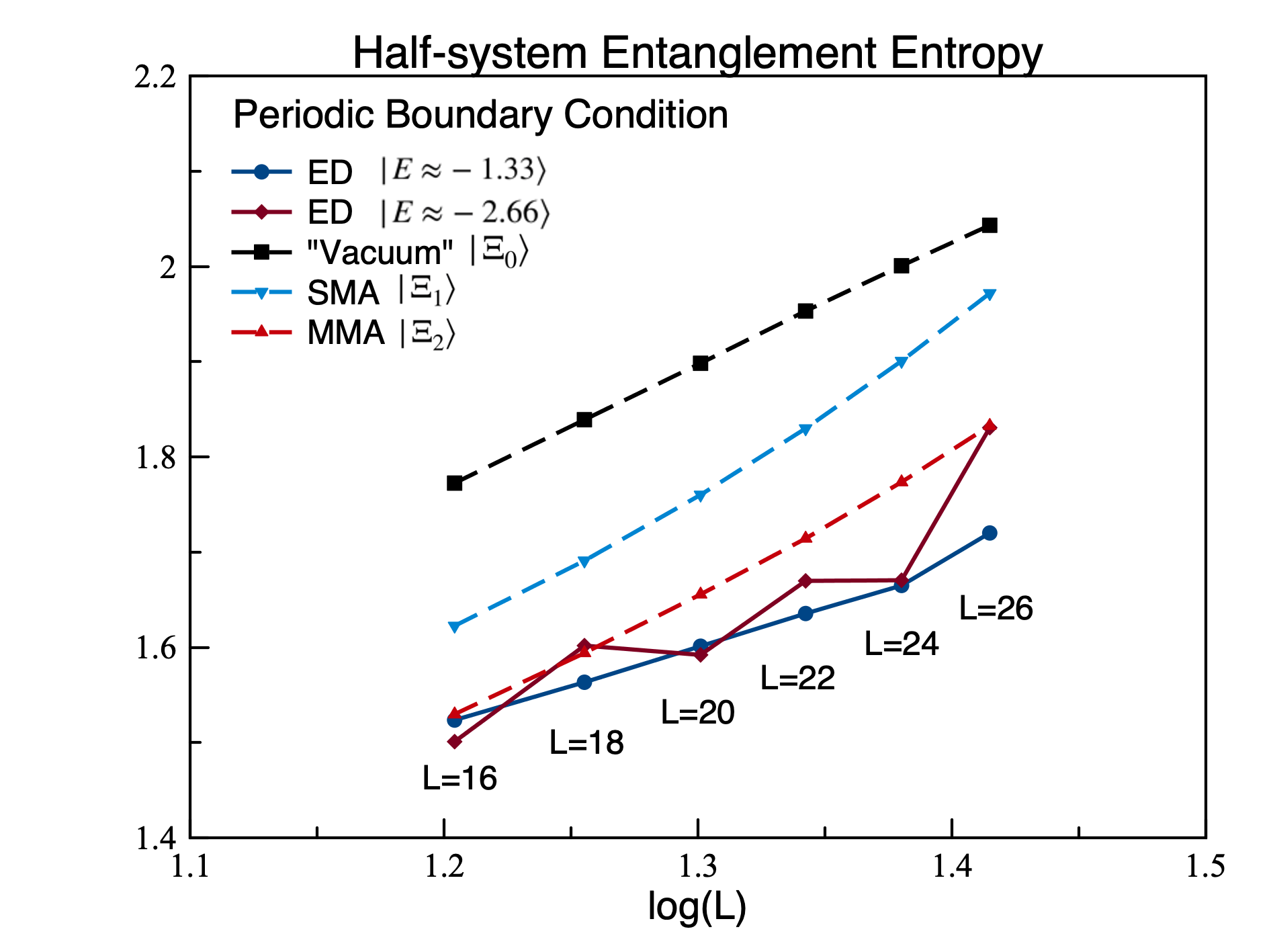}
    \caption{Bipartite entanglement entropies of the exact eigenstates and the variational SMA and MMA states.
    The primary scar states with $E \approx -1.33$ and $E \approx -2.66$ show 
    what appears to be logarithmic scaling for the available system sizes $L$. 
    The ``vacuum" $|\Xi_0 \ra \equiv [|\Phi_1 \ra + (-1)^{L_b} |\Phi_2 \ra]/\xi_0$ (i.e., exact $E = 0$ wavefunction with appropriate momentum) and the SMA/MMA $|\Xi_1 \ra$ and $|\Xi_2 \ra$ are expected to have constant entanglement scaling at large $L$, while they also show apparent
    logarithmic scaling at these small system sizes.
    Note that for the optimal SMA and MMA we found, adding quasiparticles in fact decreases the entnaglement entropies.
    }
    \label{fig:EEscaling}
\end{figure}

In this section, we discuss the bipartite entanglement entropy scaling of the SMA and MMA wavefunctions comparing with the ED results.
In Fig.~\ref{fig:EEscaling}, we show the bipartite entanglement entropy at small system sizes reachable by ED, obtained for chains in PBC for dividing the system into halves.
For the primary scar states at $E \approx -1.33$ and $E \approx -2.66$, their entanglement entropies show seeming logarithmic scaling, and are conjectured in Ref.~\cite{turnerQuantum2018} to have such scaling in the thermodynamic limit.
On the other hand, the ``vacuum" wavefunction of the quasiparticles, $|\Xi_0 \ra \equiv [|\Phi_1 \ra + (-1)^{L_b} |\Phi_2 \ra]/\xi_0$ (i.e., properly normalized exact $E=0$ eigenstate with appropriate $0$ or $\pi$ momentum depending on $L_b=L/2$),
has constant entanglement scaling in the thermodynamic limit with the saturation value $S \approx 2.254$, see the main text for details.
Since they add only a finite number of quasiparticles, the SMA and MMA wavefunctions $|\Xi_1 \ra$ and $|\Xi_2 \ra$ are also expected to have constant entanglement scaling in the thermodynamic limit. 
However, as we can see from Fig.~\ref{fig:EEscaling}, the vacuum $|\Xi_0 \ra$ and the SMA and MMA $|\Xi_{1,2} \ra$ wavefunctions also show apparent logarithmic entanglement scaling and bound the entanglement of the primary scar states at the available small system sizes.
It is also noteworthy that in our SMA and MMA wavefunctions, adding quasiparticles in fact decreases the entanglement entropy; this is contrary to common intuition about adding quasiparticles on top of a ground state, but it can happen in formal MPS states and depends on the properties of the ``excitation'' matrices.

If the SMA and MMA wavefunctions are qualitatively true asymptotic descriptions for the primary scar states, then the seeming logarithmic entanglement scaling of the ED results could be simply finite-size effect, and to see such constant scaling behavior, one may need to go to much larger system sizes.
However, while the statements about our exact $E=0$ scar states are exact, the SMA and MMA wavefunctions are only approximations to the ED scars.
One needs to study if it is possible to construct convergent improvements of the SMA and MMA states and their true properties in the thermodynamic limit, which is a non-trivial question given the surrounding eigenstates forming apparently thermal background.
We hope that addressing this question will help understanding stability of the scar states in the thermodynamic limit and in the presence of generic perturbations, while the presented entanglement data is meant to show that the available system sizes are still not sufficient to distinguish between constant or logarithmic entanglement scaling in the primary scar states.

\section{Diagonalizing the Hamiltonian in the variational space spanned by $|\Xi_n \ra$}
While increasing the number of variational parameters is one way to improve the ansatzes, we can also improve the trial states starting with $|\Xi_n \ra$ in the same spirit as the FSA improves on the states constructed using $(H^+)^n |Z_2 \ra$. 
That is, we can treat the span of $|\Xi_n \ra$, $n = 1, \dots, L_b$ as the ``variational subspace" and project the Hamiltonian into this variational space (recall that $L_b = L/2$, and here $n$ runs over the negative-energy MMA states).
More specifically, we obtain an $L_b \times L_b$ effective Hamiltonian $H_\text{eff}$ with matrix elements $[H_\text{eff}]_{nm} = \la \Xi_n|H|\Xi_m \ra$ and the overlap matrix $B$ with matrix elements $[B]_{nm} = \la \Xi_n|\Xi_m \ra$.
(Note that these matrices in fact are in block-diagonal form due to the symmetries.)
We then solve the generalized eigenvalue problem $H_\text{eff} \vec{v}^{(i)} = \lambda_i B \vec{v}^{(i)}$, obtaining the improved wavefunctions $\sum_{n=1}^{L_b} \vec{v}_n^{(i)} |\Xi_n \ra$, for $i = 1, \dots, L_b$.

\begin{figure}
    \includegraphics[width=0.5\columnwidth]{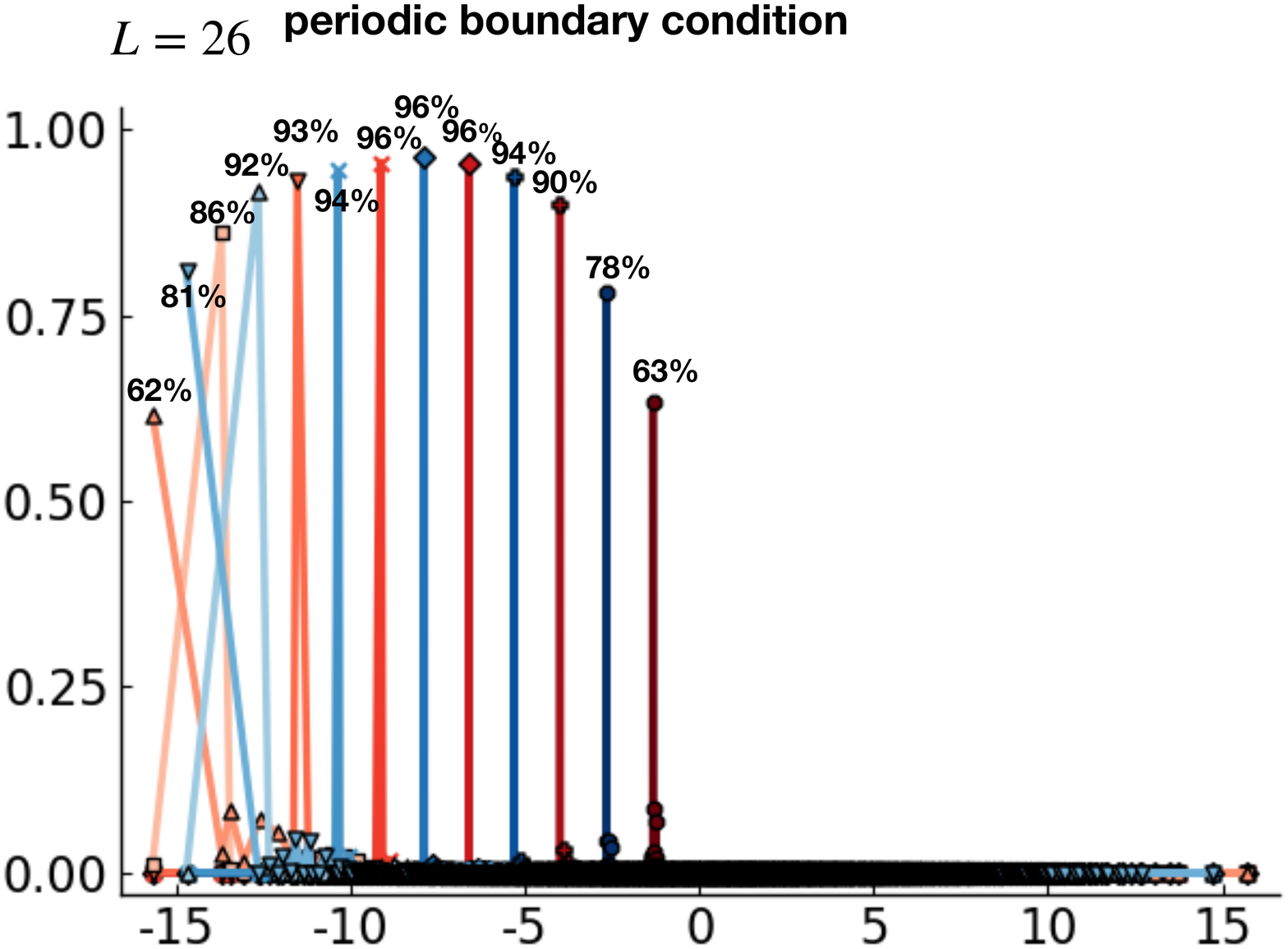}
    \caption{Improving the ``bond-dimension 2" multiparticle wavefunction by diagonalizing the projected Hamiltonian in the ``variational subspace" $\{|\Xi_n\ra|,n=1 \dots 13 \}$. 
    Such procedure improves the approximations on the ground states and scar states near the ground state.}
    \label{fig:qprediag}
\end{figure}

Figure~\ref{fig:qprediag} shows the overlap between the improved trial states and the eigenstates at $L = 26$.
We see that the improvements are mainly on the approximations on the scar states close to the ground state and the ground state; while the approximations to the  scar states close to the middle of the spectrum are not affected much.
This is expected, since, as one can see from a careful inspection of Fig.~2 in the main text, $|\Xi_n \ra$'s with $n \gtrsim L_b/2$ have high weights on usually two primary scar states.
The diagonalization procedure within this variational subspace therefore can be better isolated and improve approximations to the corresponding scar states.

To conclude, we see that, qualitatively, the primary scar states can be well understood as free quasiparticles, at least for our finite system sizes. 
An immediate question is if such a description survives for much larger sizes or even in the thermodynamic limit. We already see that some systematic improvements of the approximations can be achieved by increasing the number of variational parameters, as in the bond-dimension-3 SMA or allowing superpositions of the MMA states as in the present section.
Some immediate improvements could be achieved also by allowing the variational parameters to vary in each individual MMA state rather than simply using the values from the optimal SMA state, and by allowing superpositions of different families of the already constructed states, such as the SMA $|\tilde{\Xi}_1 \ra$ and the MMA $|\Xi_2 \ra$ for the $E \approx -2.66$ scar states, etc.
A more systematic approach is to increase the excitation block size and study convergence to the exact scar states.
In particular, we hope that this can tell whether the scar states truly survive in the thermodynamic limit even when they do not have exact closed-form expressions as happens in more fine-tuned models.
This is left for future work.

%